\documentclass{jfm}

\usepackage{graphicx}
\usepackage{epstopdf, epsfig}

\usepackage{color}
\usepackage{mathtools}
\usepackage{tabls}
\usepackage{bm}
\usepackage{bbm}

\usepackage{tikz}
\usetikzlibrary{matrix}
\usetikzlibrary{calc}
\usetikzlibrary{patterns,fadings}
\usetikzlibrary{positioning}
\usetikzlibrary{decorations.markings}
\usetikzlibrary{fadings}

\usetikzlibrary{arrows}

\newcommand\e{{\bf{e}}}

\newcommand\uu{{\bf{u}}}

\newcommand\x{{\bf{x}}}

\newcommand\AAA{{\mathsf{A}}}

\newcommand\Pec{\mbox{\textit{Pe}}}  

\newcommand\DDt{\ensuremath{\frac{d}{dt}}}
\newcommand\ddt{\ensuremath{\frac{\partial}{\partial t}}}

\definecolor{indi}{rgb}{0,0.255,0.4157}
\definecolor{deepblue}{rgb}{0,0,0.5}
\definecolor{deepred}{rgb}{0.6,0,0}
\definecolor{deepgreen}{rgb}{0,0.5,0}
\definecolor{lightblue}{rgb}{0.95,0.95,1}
\definecolor{lightgrey}{rgb}{0.6,0.6,0.6}
\usepackage{listings}
\usetikzlibrary{decorations.pathreplacing}

\shorttitle{Enhanced fluid mixing by optimisation}
\shortauthor{M.F. Eggl and P.J. Schmid}

\title{Mixing enhancement in binary fluids using optimised stirring strategies}

\author{M.F. Eggl and Peter J. Schmid} 

\affiliation{Department of Mathematics, Imperial College
  London, London SW7 2AZ, United Kingdom}

\begin{document}

\maketitle

\begin{abstract}
  Mixing of binary fluids by moving stirrers is a commonplace process
  in many industrial applications, where even modest improvements in
  mixing efficiency could translate into considerable power savings or
  enhanced product quality. We propose a gradient-based nonlinear
  optimisation scheme to minimise the mix-norm of a passive
  scalar. The velocities of two cylindrical stirrers, moving on
  concentric circular paths inside a circular container, represent the
  control variables, and an iterative direct-adjoint algorithm is
  employed to arrive at enhanced mixing results. The associated
  stirring protocol is characterised by a complex interplay of
  vortical structures, generated and promoted by the stirrers'
  action. Full convergence of the optimisation process requires
  constraints that penalise the acceleration of the moving
  bodies. Under these conditions, considerable mixing enhancement can
  be accomplished, even though an optimum cannot be guaranteed due to
  the non-convex nature of the optimisation problem. Various
  challenges and extensions of our approach are discussed.
\end{abstract}


\section{Introduction \label{sec:intro}}

The mixing of binary fluids -- the process by which a heterogeneous
mixture of two miscible fluids is manipulated into a homogeneous blend
of uniform composition -- is at the core of a great many industrial
and technological applications. The food and beverage industry, as
well as the consumer product industry abound with examples where
multiple fluid components are mixed into a final product. Adhesives,
sealants, cosmetics, inks and paints all consist of multiple
ingredients that need to be mixed into their final state during a
complex industrial process. Efficiency and consistency are paramount
in maintaining a quality product that is cost-effective to
manufacture. Some of the strictest tolerances in mixing quality can be
found in the pharmaceutical industry where medication has to be mixed
into precise doses. But also chemical engineering processes, such as
polymer production, rely on accurate mixing to facilitate the proper
chemical reactions and to reduce undesirable by-products (for an
overview of theoretical and practical aspects of mixing,
see~\cite{Handbook2003} or~\cite{Uhl2012}).

Mixing processes can be induced actively or passively. The active
strategy is commonly based on a stirrer system, where paddles or rods
agitate the binary mixture, induce vortical fluid structures and
ultimately blend the initial ingredients. The geometry, path and speed
of the stirrers have a great influence on the effectiveness and
efficiency of the mixing process and are the subject of mixing
optimisation. Passive systems, on the other hand, possess no moving
parts, but instead rely on a complex baffle system inside an
inflow-outflow device that mixes initially separated fluid
components.

In this article, we will concentrate on an active stirrer system and
develop a mathematical and computational framework for the formulation
and solution of a constrained optimisation problem that yields
favorable stirrer protocols for enhanced mixing results in binary
fluid systems. Constraints stem from speed and path restrictions on
the stirrers: stirrer systems are subject to mechanical and material
limitations, and paddles or rods often cannot accelerate or change
directions at will or too abruptly. In addition, while a significant
part of industrial mixing processes involve non-Newtonian fluids, we
will, for simplicity's sake, focus on Newtonian fluids. Furthermore,
we will concentrate on inertial, laminar mixing. The inertial aspect
of this parameter regime, described by a Reynolds number above the
Stokes-flow regime, guarantees a rich and varied control space, taking
advantage of advective, unsteady and diffusive processes, while the
laminar aspect avoids divergences of the direct-adjoint optimisation
scheme due to the existence of positive Lyapunov exponents linked
to turbulent fluid motion. Despite these restrictions, a great many
mixing processes fall into our chosen parameter regime.

Research in mixing has a long and remarkable history, covering
theoretical aspects as well as technological applications. A large
body of literature has been devoted to mixing in simplified fluid
models, for example neglecting viscosity, surface tension, density
differences or fluid inertia. The primary mechanism has been
identified as streamline stretching~\citep{Spencer1951} where the
interface between two fluids is repeatedly distorted and redistributed
into the bulk of the mixing volume. Among these simplifications,
Stokes mixing, i.e., the mixing of highly viscous fluids where
inertial effects can be neglected, has arguably received the most
attention. This tendency has been further fueled by the rise of
micro-mixers where multiple fluid components are injected into a
micro-device and extracted affter the mixing process is completed
(see, e.g., \cite{Orsi2013,Galletti2015}).

More mathematical investigations studied the breakdown in scales and
the statistical properties of the observed cascade of fluid
filaments. Iterated maps have often been used to determine measures
that describe the pertinent scale dynamics or to design optimal mixing
results in these
measures~\citep{Mathew2007,Gubanov2010,Lin2011,Finn2011}. Many of these
findings can be found in~\cite{Sturman2006}. Of particular interest
was the rise of chaotic mixing motion from pure advection, even for
laminar flows~\citep{Aref1984,Ottino1989,Liu2008}.

Rather recently, the inertial, but laminar mixing regime has been
explored using advances in optimisation techniques. These studies
build on the definition of proper mixing
measures~\citep{Mathew2005,Thiffeault2012} and break with the focus on
hydrodynamic instabilities~\citep{Balogh2005} to increase
mixing. Using wall-mounted blowing/suction control in a channel,
improved mixing could be accomplished by directly targeting a mixing
measure, rather than a flow instability~\citep{Foures2014}. Further
studies~\citep{Vermach2018,Marcotte2018} have extended this approach
to higher dimensions and stratified flows.

The present article will remain in the inertial, but laminar regime
and accomplish mixing of a binary fluid by embedded stirrers. These
stirrers are constrained to specific paths, but can move along them in
a manner that enhances mixing over a user-specified time horizon. The
mixedness of the binary fluid is quantified using
mix-norms~\citep{Mathew2007,Thiffeault2012} for a passive scalar; an
objective based on this measure is then optimised by a nonlinear,
gradient-based scheme, which in turn provides the associated stirring
protocol.

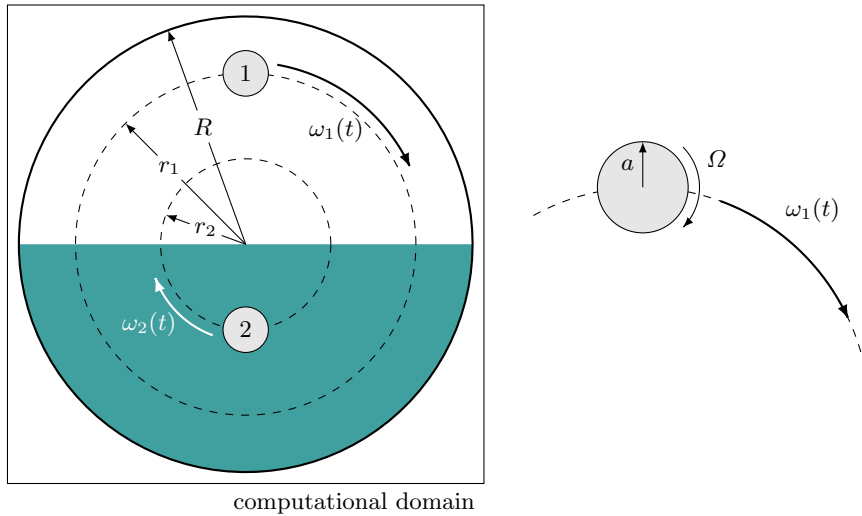
\begin{figure}
  \centering
  \begin{tikzpicture}[scale=1.5,>=latex]
    \draw[black,thin] (-2.1,-2.1) rectangle (2.1,2.1);
    \fill[teal!75, domain=180:360, samples=100] plot ({2*cos(\x)}, {2*sin(\x)});
    \draw[black, thick, domain=0:360, samples=100] plot ({2*cos(\x)}, {2*sin(\x)});
    \draw[black,dashed] (0,0) circle [radius=1.5cm];
    \fill[gray!20] (0,1.5) circle [radius=0.2cm];
    \draw[black] (0,1.5) circle [radius=0.2cm];
    \draw[black,dashed] (0,0) circle [radius=0.75cm];
    \fill[gray!20] (0,-0.75) circle [radius=0.2cm];
    \draw[black] (0,-0.75) circle [radius=0.2cm];
    \draw[black, thick, <-, domain=25:80, samples=100] plot ({1.6*cos(\x)}, {1.6*sin(\x)});
    \draw[white, thick, <-, domain=200:250, samples=100] plot ({0.85*cos(\x)}, {0.85*sin(\x)});
    \draw[black,thin] (0,0) -- ({0.95*cos(110)},{0.95*sin(110)});
    \draw[black,thin,->] ({1.25*cos(110)},{1.25*sin(110)}) -- ({2*cos(110)},{2*sin(110)});
    \node at ({1.1*cos(110)},{1.1*sin(110)}) {$R$};

    \draw[black,thin] (0,0) -- ({0.8*cos(135)},{0.8*sin(135)});
    \draw[black,thin,->] ({1.1*cos(135)},{1.1*sin(135)}) -- ({1.5*cos(135)},{1.5*sin(135)});
    \node at ({0.95*cos(135)},{0.95*sin(135)}) {$r_1$};

    \draw[black,thin] (0,0) -- ({0.225*cos(160)},{0.225*sin(160)});
    \draw[black,thin,->] ({0.525*cos(160)},{0.525*sin(160)}) -- ({0.75*cos(160)},{0.75*sin(160)});
    \node at ({0.375*cos(160)},{0.375*sin(160)}) {$r_2$};
    \node at (0.8,1) {$\omega_1(t)$};
    \node[white] at (-0.85,-0.7) {$\omega_2(t)$};
    \node at (0,1.5) {$1$};
    \node at (0,-0.75) {$2$};
    \node[anchor=north east] at (2.1,-2.1) {\footnotesize{computational domain}};

    \def\s{3.5};
    \def\t{-0.5};
    \draw[black, thin, dashed, domain=0:120, samples=150] plot ({\s+2*cos(\x)}, {\t-1+2*sin(\x)});
    \draw[black, thick, domain=25:70, samples=100,<-] plot ({\s+2*cos(\x)}, {\t-1+2*sin(\x)});
    \fill[gray!20] (\s,\t-1+2) circle [radius=0.4cm];
    \draw[black] (\s,\t-1+2) circle [radius=0.4cm];
    \draw[black, domain=-45:45, samples=100,<-] plot ({\s+0.5*cos(\x)}, {\t-1+2+0.5*sin(\x)});
    \draw[black, thin,->] (\s,\t+1) -- (\s,\t+1.5-0.1);

    \node at (\s+1.5,\t+0.8) {$\omega_1(t)$};
    \node at (\s+0.65,\t+1.25) {$\Omega$};
    \node[anchor=east] at (\s,\t+1.2) {$a$};

  \end{tikzpicture}
  \caption{\label{fig:MixSetup} Setup of mixing problem. A circular
    domain of radius $R$ encloses two stirrers of circular
    cross-section (given by radius $a$) on concentric circular paths
    of fixed radii $r_{1,2}.$ The stirring strategy is determined by
    the time-dependent circular velocities $\omega_{1,2}(t),$ together
    with the rotation rate $\Omega$ about each cylinder's
    axis. Initially, one fluid component occupies the lower half of
    the vessel, while the other fluid component resides in the upper
    half.}
\end{figure}

\section{Mathematical framework \label{sec:math}}

\subsection{Governing equations}

The focus of this study is the mixing process of a binary, miscible
and Newtonian fluid by multiple circular stirrers on prescribed paths,
and its optimisation by manipulating the stirring strategy within
specified constraints. A two-dimensional configuration is considered.
The process can be simulated by solving the fluid equations of motion,
augmented by a transport equation for a passive scalar $\theta.$ We
have
\begin{subeqnarray}
 \ddt \uu + \uu \cdot \nabla \uu + \frac{1}{C_\eta}\left(\chi \uu -
 \chi_k \uu_{s,k} \right) + \nabla p - \frac{1}{\Rey} \nabla^2 \uu &=&
 0, \\
 \nabla \cdot \uu &=& 0, \\
 \ddt \theta + \left( (1-\chi) \uu + \chi_k \uu_{s,k} \right) \cdot
 \nabla \theta - \nabla \cdot \left( \left[ \frac{1}{\Pec} (1-\chi) +
   \frac{\chi}{C_{\eta}} \right] \nabla \theta \right) &=& 0.
  \label{eq:GovEqu}
\end{subeqnarray}
with $\uu$ as the velocity vector, $p$ as the pressure field, and
$\theta$ as a passive scalar (ranging from zero in one fluid to one in
the other). The governing equations have been expressed in
non-dimensional form using a characteristic length $L_0$ and velocity
scale $u_0.$ This choice introduces the Reynolds number $\Rey$ and the
P{\'e}clet number $\Pec,$ to express the kinematic viscosity and the
diffusion coefficient of the mixing fluid in non-dimensional
form. Furthermore, the system of equations~(\ref{eq:GovEqu}) contains
terms that model the embedded stirrers via a Brinkmann penalisation
approach~\citep[see][]{Angot1999}. The multiple solid stirrers, indexed by
the subscript $k,$ are characterised by their velocity $\uu_{s,k},$
and are taken as circular in cross-section. The masks $\chi_k$
describing the embedded solid bodies equal one for points occupied by
the $k$-th stirrer and zero outside of it. The mask $\chi$ accounts
for the overall geometry, such as the domain boundaries. The constant
$C_\eta$ ensures the rapid relaxation of the fluid variables towards
the respective values imposed by the stirrers or the geometry. The
above formalism allows the efficient treatment of objects moving
through a background grid on which the motion of the surrounding fluid
is describes. Details of this approach and its numerical
implementation can be found in~\cite{EgglSchmid2018}. The setup shown
in~(\ref{eq:GovEqu}) imposes no-slip velocity boundary conditions on
the stirrers and Neumann conditions for the passive scalar on the
solid bodies.

\subsection{Measuring mixedness}

In anticipation of our stated goal of enhancing mixing efficiency, we
have to introduce a measure that quantifies the degree of mixedness of
a particular flow state. This measure shall be based solely on the
passive scalar field $\theta.$

In general, mixing is defined as the reduction of inhomogeneities of a
given indicator field~\citep{Handbook2003}, which still leaves open a
precise mathematical definition to be used in our case. Several norms
of the passive scalar $\theta$ that attempt to mathematically define
the measure of mixedness have been proposed and used in the
past~\citep{Mathew2005}, among them the variance or the more complex
negative-index and fractional-index Sobolev
norms~\citep{Thiffeault2012,Foures2014}. As the choice of norm may
influence the outcome of the optimisation, but will not affect the
design of our computational optimisation platform, we will focus on
the Sobolev norm of negative fractional index of the passive scalar
$\theta.$ A measure of this general type downplays the role of small
scales and instead directs mixing efforts towards larger fluid
elements. It attains higher values for an unmixed field (with high
levels of inhomogeneities) and decreases as the scalar field becomes
more mixed. Mathematically, the mixedness measure is given as

\begin{equation}
  \Vert \theta \Vert_{\rm{mix}} \equiv \frac{1}{\vert \Omega
    \vert}\int_{\Omega} \Vert \nabla^{-2/3} \theta(\x,t) \Vert \text{
    d}\Omega,
    \label{eq:SobolevNorm}
\end{equation}
with $\Omega$ denoting our computational domain, and $\vert \Omega
\vert$ representing its size (volume or area). In the above
definition, we have assumed, without loss of generality, a zero mean
of the passive scalar field $\theta.$ Throughout this paper we will be
optimising with respect to this quantity, but we stress again that
other norms can be employed without conceptual changes in the
optimisation procedures. The fractional exponent of $2/3$ can be
justified using arguments from optimal transport and ergodic theory.
Examples of previous studies using mix-norm optimisation employed
$-1/2$ \citep{Foures2014} or $-1$ \citep{Lin2011}.

\subsection{Mixing protocol}

As a first attempt at optimising the mixing of a binary fluid, we will
concentrate on a stationary circular vessel with two stirrers on
circular paths or distinct radii (see figure~\ref{fig:MixSetup} for a
sketch of this configuration). The stirrers have a circular
cross-section, and their velocities along their respective paths are
undetermined and subject to optimisation and constraints. The circular
path is conveniently defined in polar coordinates, while we formulate
the remaining equations in Cartesian coordinates, and we thus
introduce the vector-valued function $\bm{l}$, which transforms
between the two coordinate systems according to $(l_1(\phi),l_2(\phi))
= (-\sin(\phi),\cos(\phi))$ with $\phi$ as the angle traversed along
the path of the circle. The parameterisation of the velocity of the
$k$-th stirrer thus becomes
\begin{eqnarray}
\bm{u}_{s,k} &=&\omega_{k}(t)\bm{r}_k(\bm{x})\bm{l}(\varphi_k(t)) + \Omega_k
     {\bf{a}}_k, \label{FLUSI:US}
\end{eqnarray}
where $\varphi$ is the sum of the angles travelled along the path, i.e.,
\begin{align}
    \varphi_k(t) = \int_0^t\omega_{k}(s) \text{d}s.
\end{align}
Following the notation of the governing equations, we define $\omega_k(t)$
as the rotational speed of the $k$-th solid about the
centre of the vessel, ${\bf{r}}_k$ denotes the distance from the same
centre, $\Omega_k$ stands for the rotation about the stirrer's centre
and ${\bf{a}}_k$ represents the stirrer's (vectorial) radius. For
simplicity, we take $\Omega_k = 0;$ the alternative choice $\Omega_k =
\omega_k$ yielded very similar results in enhancing mixing efficiency.

\subsection{Constrained optimisation}

We can then state the optimisation problems as finding a
time-dependent velocity protocol $\omega_k(t)$ for each of the two
stirrers such that the mix-norm of the passive scalar is minimised
over a prescribed time horizon. This minimum has to be achieved while
satisfying the governing equations and respecting constraints and
bounds on the stirrer velocities. Mathematically we have

\begin{subeqnarray}
  && \min \left\{ \int_0^T \Vert \theta \Vert_{\rm{mix}} \ dt \right\}
  \\
  && \hbox{subject to equations}~(\ref{eq:GovEqu}) \\
  && \hbox{ and} \quad \int_0^T \sum_k \Vert \uu_{s,k} \Vert^2 \ dt
  \leq E_0 \\
  && \hbox{ and} \quad \uu_{s,\rm{lower}} \leq \uu_{s,k} \leq
  \uu_{s,\rm{upper}} \qquad k=1,2, \\
      && \hbox{ and} \quad \bm{a}_{s,\rm{lower}} \leq \bm{a}_{s,k} \leq
  \bm{a}_{s,\rm{upper}} \qquad k=1,2.
  \label{eq:constropt}
\end{subeqnarray}
The constraints on the stirrer strategy are threefold: the first
constraint limits the $L_2$-norm of $\bm{u}_s$, i.e., the kinetic
energy of stirrers' motion along their paths, expended over the time
horizon $T$ to a maximum value of $E_0;$ the second and third impose
upper and lower bounds directly on the stirrers' velocities and
accelerations, respectively. All restrictions could conceivably stem
from mechanical limitations of the mixing apparatus. In our study, we
will consider the constraints successively in order to determine the
influence they impose on the optimisation results.

\section{Computational framework \label{sec:comp}}

The implementation of the above optimisation problem requires the
discretisation of the governing equations and the reformulation of the
constrained problem~(\ref{eq:constropt}) in terms of an unconstrained
one.

\subsection{Numerical scheme for the governing equations}

Starting point for the numerical treatment of mixing enhancement is
the open-source FLUSI software~\citep{Engels2015}, in which the
governing equations are discretised on a Cartesian, double-periodic
domain for the two-dimensional case. This formulation allows the
application of Fourier-spectral techniques to represent the spatial
derivatives. The outer perimeter of the mixing vessel (with radius
$R$) and the two stirrers on their respective circular paths (with
radius $r_1$ and $r_2$) are described by a Brinkman penalisation
technique as shown in~(\ref{eq:GovEqu}). The original software has
been augmented by the passive scalar field and embedded into a
gradient-based optimisation formalism.

The Fourier-spectral discretisation allows the replacement of spatial
derivatives with a multiplication by components of a wavenumber vector
${\bf{k}} = \left({\bf{k}}_1,{\bf{k}}_2\right)^T,$ with the indices
${}_{1,2}$ indicating the two coordinate directions. Mathematically,
we introduce this replacement as $\partial/\partial x_j \to \AAA_j,$
with $\AAA_j = \hbox{diag}\{ i{\bf{k}}_j \}.$ The semi-discretised set
of equations then reads

\begin{subeqnarray}
  \DDt \uu_i + \uu_j \circ \left[ \AAA_j \uu_i \right] +
  \frac{1}{C_\eta}\left(\chi \circ \uu_i - \chi_k \circ
  {\uu_{s,k}}_i\right) + \AAA_i p - \frac{1}{Re} \AAA_j \AAA_j \uu_i &=&
  0, \\
  \AAA_j \uu_j &=& 0, \\
  \DDt \theta + \left( ({\bf{1}} - \chi) \circ \uu_j + \chi_k \circ
       {\uu_{s,k}}_j \right) \circ \left[ \AAA_j \theta \right] -
       \AAA_j \left[ \frac{1}{\Pec} \left( {\bf{1}} - \chi \right) +
         \kappa \chi \right] \circ \AAA_j \theta &=& 0 \phantom{123456}
\end{subeqnarray}
where we introduced the Hadamard (element-wise) product $\circ$
(see~\cite{Horn2012}) and assumed implicit (Einstein) summation over
identical indices.

Particular care has to be exercised when evaluating the nonlinear
terms, as aliasing errors can lead to inaccuracies and numerical
instabilities. A low-pass Hou-Li filter~\citep{HouLi2007} has been
applied to avoid scattering of unresolved, small scales onto resolved,
large scales. In addition, P3DFFT \citep{pekurovsky2012}, a highly
efficient, parallel Fourier-transform library, is used to ensure
scaling on parallel computer architectures.

Finally, the representation of solid bodies on an underlying Cartesian
grid calls for a transfer of geometric information onto the background
mesh. This transfer is accomplished by mollified delta-functions,
smoothing the otherwise discontinuous mask onto the grid and thus
avoiding numerical inaccuracies and
instabilities~\citep{Kolomenskiy2009}.

\subsection{From constrained to unconstrained optimisation}

A common reformulation of the constrained optimisation
problem~(\ref{eq:constropt}) as an unconstrained problem introduces
Lagrange multipliers (or adjoint variables) for the dependent
variables of equations~(\ref{eq:GovEqu}): the adjoint velocity will
enforce the momentum equation, the adjoint pressure the divergence
condition, and the adjoint passive scalar the transport equation for
$\theta.$ The augmented Lagrangian -- consisting of the cost
functional and the scalar product of the adjoint variables and the
governing equations -- then needs to minimised. A system of equations,
referred to as the KKT-system, can then be derived by setting to zero
the first variation of the augmented Lagrangian with respect to direct
(original) and adjoint variables. The first variation with respect to
the adjoint variables recovers the original set of governing
equations. The first variation with respect to the original variables
produces, after considerable algebra, a set of equations governing the
adjoint variables. The first variation with respect to the control
variables (in our case, the velocity strategies $\omega_{1,2}(t)$)
will furnish the gradients of the cost functional with respect to
$\omega_{1,2}(t)$ which will be used to enhance the mixedness of our
fluid system via improved stirring strategies.

\subsection{Adjoint equations}

Denoting the adjoint variables (velocity, pressure, passive scalar) by
$\uu^{\dag}$, $p^{\dag}$ and $\theta^{\dag},$ the governing equations
for their evolution, in semi-discretized form, read

\begin{subeqnarray}
  \DDt \uu^{\dag}_i - \Pi^{\dag}_k \circ [\AAA_i \uu_k ] -
       \AAA_j^H [\uu_j \circ
         \Pi^{\dag}_i] - \frac{\chi}{C_{\eta}}
       \circ \Pi^{\dag}_i + \frac{1}{\Rey}\AAA_j^H
             \AAA_j^H \uu_i^{\dag} && \nonumber \\
   -({\bf{1}}-\chi) \circ \theta^{\dag} \circ[ \AAA_i \theta]
   &=& 0 \\
  \AAA_j^H \Pi_j^\dag &=& 0 \\
  \DDt \theta^{\dag} - \AAA_j^H [({\bf{1}}-\chi) \circ \uu_j
    \circ \theta^{\dag}] + \AAA_i^H ([\frac{1}{\Pec}({\bf{1}}-\chi) +
    \kappa\chi] \circ \AAA_i^H \theta^{\dag}) && \nonumber \\
  -\AAA_j^H [ \chi_i \circ (\uu_{s,i})_j\ \circ \theta^{\dag}]
  &=& 0
  \label{Adjoint:EqEnd}
\end{subeqnarray}
with terminal conditions
\begin{align}
  \bm{u}^{\dag}(\bm{x},T) = 0, \qquad
  \theta^{\dag}(\bm{x},T) =  \frac{2}{V_\Omega}(\mathsf{A}_i^{-2/3})^H(\mathsf{A}_i^{-2/3}\theta(\bm{x},T)).
\end{align}
It is important to note that the above adjoint equations are linear in
the adjoint variables, but are dependent on the direct variables
$\uu_i.$ Moreover, it should become apparent that the adjoint
equations have to solved backwards in time, from $t=T$ to $t=0.$

The optimality conditions, stemming from the first variation with
respect to the control variables, result in the adjoint rotational
velocity along the circular paths given by

\begin{equation}
  \omega_{k}^\dag = r_i\left[l_j(\varphi(t))
    +\frac{\omega_{k}}{\dot{\omega_{k}}} \frac{\partial l_j}{\partial
      \varphi}\right]\chi_i^H\left(\frac{\Pi^{\dag}_j}{C_\eta}-
  (\theta^{\dag}\circ[\mathsf{A}_j\theta])\right)
\end{equation}
where $\Pi^{\dag}_i = \uu^{\dag}_i + \AAA^{H}_i p^{\dag}$. This value
provides the gradient information in our iterative optimisation scheme
and, together with a line-search routine, updates the current stirring
protocol to a more effective one. The optimisation terminates when no
more progress can be made, and the magnitude of the cost functional
gradient drops below a prescribed threshold.

\subsection{\label{sec:optim} Implementing additional constraints}

Additional constraints that need to be enforced are incorporated into
the gradient-based optimisation routine. This is accomplished by
projections and thresholding. In this case, the gradient -- computed
from the adjoint equations and the optimality condition, without
imposed constraints -- is projected and properly curtailed to comply
with energy constraints and velocity bounds. Details of the numerical
implementation of this procedure can be found in~\cite{EgglSchmid2018}.

\subsection{Summary of optimisation procedure}

The full optimisation scheme then proceeds along the following
lines. Starting with an initial guess of the stirring protocols
$\omega_{1,2}(t),$ we solve the governing equations~(\ref{eq:GovEqu})
forward in time over a chosen time horizon $[0,\ T].$ In a second
step, the adjoint set of equations~(\ref{Adjoint:EqEnd}) are solved,
starting with the proper terminal condition, backwards in time from
$t=T$ to $t=0.$ The adjoint variables are then used to evaluate the
optimality condition and retrieve the gradient of the cost functional
with respect to $\omega_{1,2}.$ This gradient is then furnished to a
standard optimisation routine (such as steepest descent or conjugate
gradients) which, together with a line-search routine, produces a new
and improved stirring protocal. In this last step, all constraints
imposed on the stirring functions will be imposed by the aforementioned projections and
thresholding. After this step, the next iteration is started. The
optimisation terminates when no further progress can be made, within
the constraints imposed on the system.

It is worth mentioning that additional complications arise from the
fact that the governing equations~(\ref{eq:GovEqu}) are nonlinear and,
as a consequence, the adjoint equations~(\ref{Adjoint:EqEnd}), while
linear in the adjoint variables, depend on the direct variables
${\bf{u}}, \theta.$ This dependency requires the storage of direct
fields during the simulation of~(\ref{eq:GovEqu}) and their injection
into~(\ref{Adjoint:EqEnd}), in reverse order, during the integration
of the adjoint system. For efficiency reasons, this exchange between
direct and adjoint simulations is handled by checkpointing, where we
trade excessive storage requirements for an increased simulation
time. The {\tt{revolve}} library~\citep{GriewankWalther2000} accomplishes
this task in an optimal manner.

\section{Test cases and results \label{sec:results}}

We follow the setup shown in figure~\ref{fig:MixSetup} with two
circular cylinders of radius $\Vert a_{1,2} \Vert = 1,$ moving on two
concentric circular paths of radius $r_1 = 3.5$ and $r_2 = 1.5$ and
embedded in a circular (stationary) vessel of radius $R = 5.$ The
Reynolds number and P\'eclet number are chosen as $Re = Pe = 1000.$

\begin{figure}
\centering
\begin{tikzpicture}[scale=1,>=latex]
  \def\e{0.1};
  \draw[black,thick] (0,0) -- (9,0) -- (9+\e,-0.2) -- (9+2*\e,0.2) -- (9+3*\e,0) -- (12,0);
  \draw[black,thick] (0,-0.2) -- (0,0.2);
  \draw[black,thick] (0.5,-0.2) -- (0.5,0.2);
  \draw[black,thick] (4,-0.2) -- (4,0.2);
  \draw[black,thick] (12,-0.2) -- (12,0.2);
  \node[anchor=west,rotate=90] at (0,2*\e) {\scriptsize{T=0}};
  \node[anchor=west,rotate=90] at (0.5,2*\e) {\scriptsize{T=1}};
  \node[anchor=west,rotate=90] at (4,2*\e) {\scriptsize{T=8}};
  \node[anchor=west,rotate=90] at (12,2*\e) {\scriptsize{T=32}};
  \draw[deepred,thick,<->] (0,-0.75) -- (0.5,-0.75);
  \draw[deepred,thick,<->] (0,-1) -- (4,-1);
  \node[anchor=west] at (4,-0.875) {\scriptsize{control strategy (actuation window)}};
  \draw[deepgreen,thick,<->] (0,-1.5) -- (4,-1.5);
  \node[anchor=west] at (4,-1.5) {\scriptsize{adjoint information (predictive horizon)}};
  \draw[deepblue,thick,<->] (0,-2) -- (9,-2) -- (9+\e,-2-0.2) -- (9+2*\e,-2+0.2) -- (9+3*\e,-2) -- (12,-2);
  \node[anchor=west] at (4,-2.25) {\scriptsize{final simulation and visualisation}};
\end{tikzpicture}
\caption{\label{fig:TimeHorizons} Sketch of time horizons for the
  optimisation problem. A control strategy (red) is applied over two
  control horizons, $T_{\rm{control}}=1$ (for short-time control) and
  $T_{\rm{control}}=8$ (for long-time control). The gradient
  information about the flow development (green), encoded in the
  adjoint variables, is gathered over a predictive horizon of
  $T_{\rm{info}}=8.$ The final simulation, based on the optimised
  strategy, is performed over $T_{\rm{sim}}=32$ non-dimensional time units
  (blue).}
\end{figure}
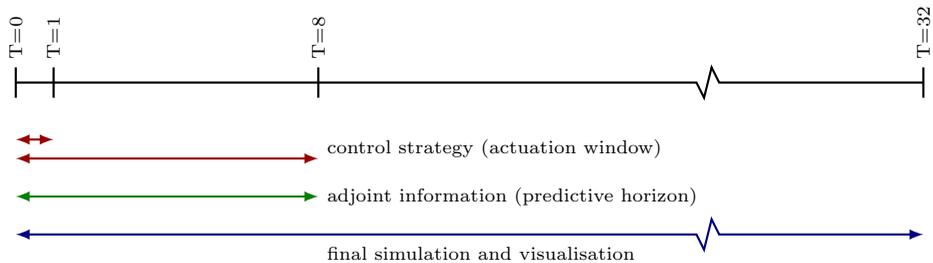

A further parameter in the optimisation concerns the time horizons
over which (i) the control strategy is applied, and (ii) over which
gradient information is gathered. The former time interval determines
the window given to the stirrers to be active mixers; after this
window is passed, the motion of the stirrers will stop, and only the
remaining inertia of the fluid and diffusion will contribute to
further mixing. The latter time interval determines the amount of
information extracted from the evolution process that is used to
compute an enhanced stirring protocol (applied over the former time
window). The control horizon may be chosen shorter than the
information (predictive) horizon: in this case, a time-compressed
strategy will be employed that accounts for, and optimises over, a
more expansive time window. In our case, we will juxtapose a
short-term strategy with $T_{\rm{control}}=1$ and a longer-time
strategy with $T_{\rm{control}}=8$ and assess the optimised strategies
in either case. Both protocols, however, have access to information
over a temporal interval of $T_{\rm{info}} = 8.$ Finally, the
simulations have been continued over $T_{\rm{sim}}=32$ units to track
the further development of the instigated mixing processes; rest
inertia and diffusion will remain the only mechanisms during this
stage. A summary of our choice of parameters is sketched in
figure~\ref{fig:TimeHorizons}.

\begin{figure}
  \centering
  \begin{tabular}{cccc}
    \rotatebox{90}{\hspace{0.5truecm} plunging} &
    \begin{tikzpicture}[>=stealth]
      \draw (0,0) node[inner sep=0] {\includegraphics[width=0.2\textwidth]{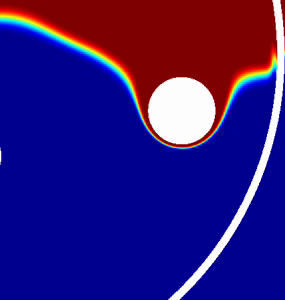}};
      \draw[gray,thick,->] (0.375,0.375) -- (0.25,0.15);
      \draw[white] (-1.1,-1.1) node {\scriptsize{(a)}};
    \end{tikzpicture} & \hspace{-0.2truecm}
    \begin{tikzpicture}[>=stealth]
      \draw (0,0) node[inner sep=0] {\includegraphics[width=0.2\textwidth]{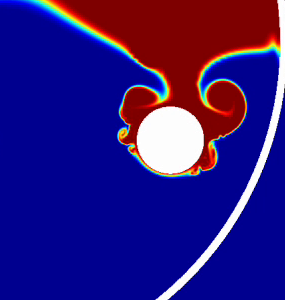}};
      \draw[gray,thick,->] (0.275,0.085) -- (0.15,-0.135);
      \draw[white] (-1.1,-1.1) node {\scriptsize{(b)}};
    \end{tikzpicture} & \hspace{-0.2truecm}
    \begin{tikzpicture}[>=stealth]
      \draw (0,0) node[inner sep=0] {\includegraphics[width=0.2\textwidth]{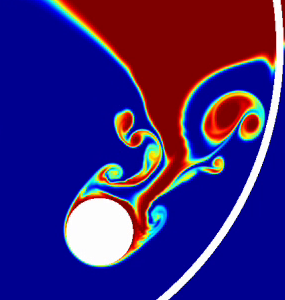}};
      \draw[gray,thick,->] (-0.4,-0.8) -- (-0.595,-0.975);
      \draw[white] (-1.1,-1.1) node {\scriptsize{(c)}};
    \end{tikzpicture} \\
    \rotatebox{90}{\hspace{0.5truecm} vortex cannon} &
    \begin{tikzpicture}[>=stealth]
      \draw (0,0) node[inner sep=0] {\includegraphics[width=0.2\textwidth]{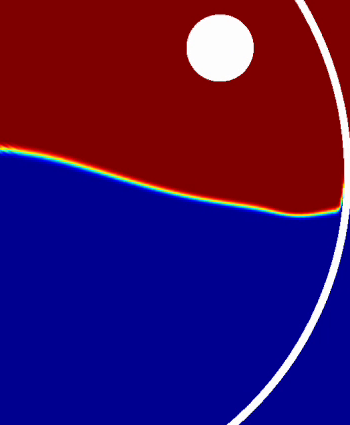}};
      \draw[gray,thick,<->] (0.28,1.475) -- (0.42,1.025);
      \fill[white] (0.35,1.25) circle [radius=0.2mm];
      \draw[white] (-1.1,-1.33) node {\scriptsize{(d)}};

      \draw[white,thick,densely dashed,->] (0.75,1) to [out=-65,in=90] (0.9,0.3);
      \draw[white,thick,densely dashed,->] (0.2,0.9) to [out=-65,in=0] (-0.1,0.4);

    \end{tikzpicture} & \hspace{-0.2truecm}
    \begin{tikzpicture}
      \draw (0,0) node[inner sep=0] {\includegraphics[width=0.2\textwidth]{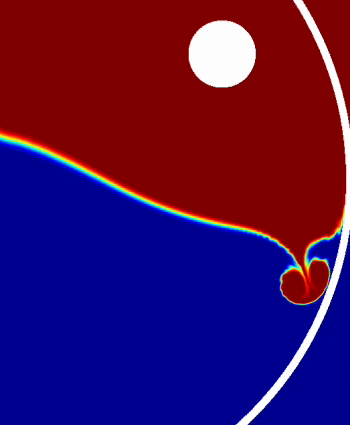}};
      \fill[gray] (0.375,1.21) circle [radius=0.35mm];
      \draw[white] (-1.1,-1.33) node {\scriptsize{(e)}};
    \end{tikzpicture} & \hspace{-0.2truecm}
    \begin{tikzpicture}
      \draw (0,0) node[inner sep=0] {\includegraphics[width=0.2\textwidth]{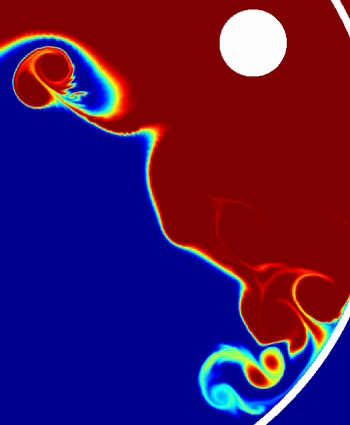}};
      \fill[gray] (0.6,1.3) circle [radius=0.35mm];
      \draw[white] (-1.1,-1.33) node {\scriptsize{(f)}};
    \end{tikzpicture} \\
    \rotatebox{90}{\hspace{0truecm} vortex collision} &
    \begin{tikzpicture}[>=stealth]
      \draw (0,0) node[inner sep=0] {\includegraphics[width=0.2\textwidth]{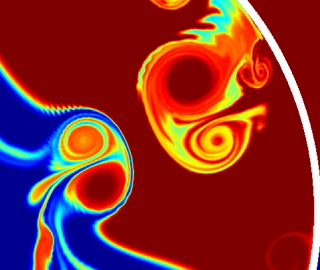}};
      \draw[white,thick,densely dashed,->] (-0.4,-0.2) -- (-0.15,-0.1);
      \draw[white,thick,densely dashed,<-] (-0.15,-0.1) -- (0.1,0);
      \draw[white] (-1.1,-0.9) node {\scriptsize{(g)}};
    \end{tikzpicture} & \hspace{-0.2truecm}
    \begin{tikzpicture}
      \draw (0,0) node[inner sep=0] {\includegraphics[width=0.2\textwidth]{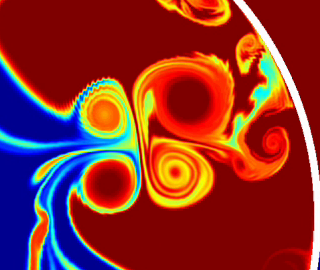}};
      \draw[white] (-1.1,-0.9) node {\scriptsize{(h)}};
    \end{tikzpicture} & \hspace{-0.2truecm}
    \begin{tikzpicture}
      \draw (0,0) node[inner sep=0] {\includegraphics[width=0.2\textwidth]{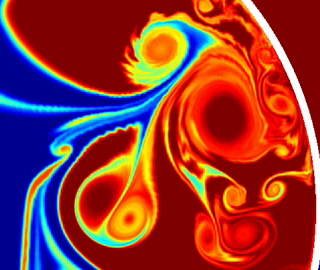}};
      \draw[white] (-1.1,-0.9) node {\scriptsize{(i)}};
    \end{tikzpicture} \\
    \rotatebox{90}{\hspace{1truecm} wall collision} &
    \begin{tikzpicture}[>=stealth]
      \draw (0,0) node[inner sep=0] {\includegraphics[width=0.2\textwidth]{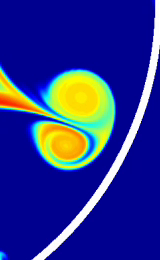}};
      \draw[white,thick,densely dashed,->] (-0.4,-0.2) -- (-0.15,-0.1);
      \draw[white,thick,densely dashed,<-] (-0.15,-0.1) -- (0.1,0);
      \draw[white] (-1.1,-1.9) node {\scriptsize{(j)}};
    \end{tikzpicture} & \hspace{-0.2truecm}
    \begin{tikzpicture}
      \draw (0,0) node[inner sep=0] {\includegraphics[width=0.2\textwidth]{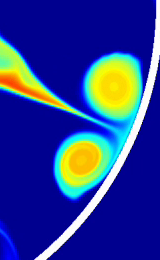}};
      \draw[white] (-1.1,-1.9) node {\scriptsize{(k)}};
    \end{tikzpicture} & \hspace{-0.2truecm}
    \begin{tikzpicture}
      \draw (0,0) node[inner sep=0] {\includegraphics[width=0.2\textwidth]{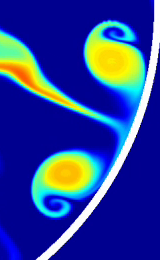}};
      \draw[white] (-1.1,-1.9) node {\scriptsize{(l)}};
    \end{tikzpicture} \\
    \rotatebox{90}{\hspace{0.33truecm} obstruction} &
    \begin{tikzpicture}[>=stealth]
      \draw (0,0) node[inner sep=0] {\includegraphics[width=0.2\textwidth]{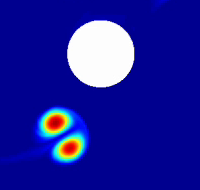}};
      \fill[gray] (0,0.55) circle [radius=0.35mm];
      \draw[white,thick,densely dashed,->] (-0.3,-0.45) to [out=30,in=230] (-0.1,-0.3) to [out=50,in=265] (0,0);
      \draw[white] (-1.1,-1.05) node {\scriptsize{(m)}};
    \end{tikzpicture} & \hspace{-0.2truecm}
    \begin{tikzpicture}
      \draw (0,0) node[inner sep=0] {\includegraphics[width=0.2\textwidth]{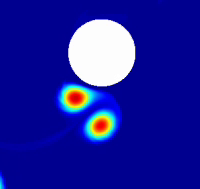}};
      \fill[gray] (0.025,0.55) circle [radius=0.35mm];
      \draw[white] (-1.1,-1.05) node {\scriptsize{(n)}};
    \end{tikzpicture} & \hspace{-0.2truecm}
    \begin{tikzpicture}
      \draw (0,0) node[inner sep=0] {\includegraphics[width=0.2\textwidth]{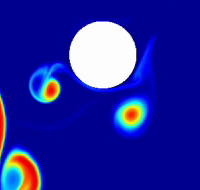}};
      \fill[gray] (0.035,0.525) circle [radius=0.35mm];
      \draw[black] (-1.1,-1.05) node {\scriptsize{(o)}};
    \end{tikzpicture}
  \end{tabular}
  \caption{\label{fig:Strategies} Various mixing strategies, from
    snapshots of the simulations. (a,b,c) Plunging of the stirrer
    through the interface, (d,e,f) casting of start-stop vortices
    towards the interface (vortex cannon), (g,h,i) collision of
    vortices, (j,k,l) collision with the vessel wall, and (m,n,o)
    breakup of vortical structures by stationary stirrers
    (obstruction).}
\end{figure}
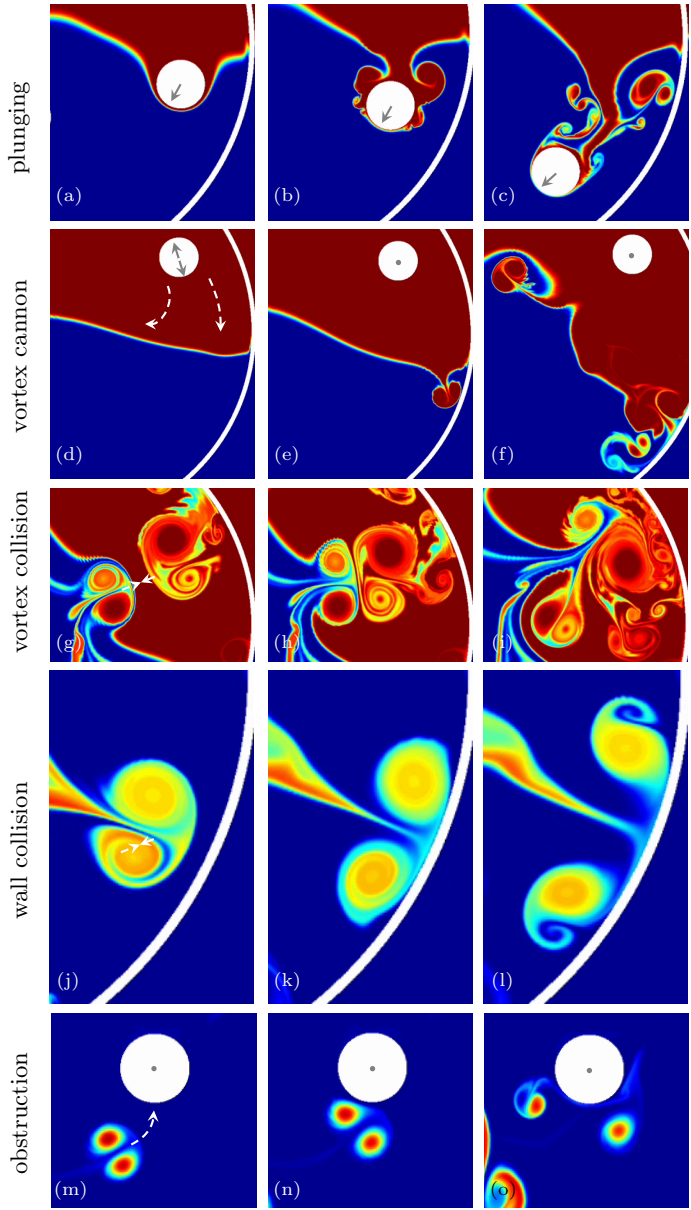

Before proceeding to the various optimisation studies, it is
instructive to reflect upon possible mixing mechanisms given the setup
in figure~\ref{fig:MixSetup}. The most obvious strategy for mixing a
binary fluid consists of a {\bf{plunging}} motion, where the cylinders
push through the initial interface, distort it and drag fluid one into
regions occupied by fluid two, and vice versa
(figures~\ref{fig:Strategies}{\it{a--c}}). This type of strategy is
nearly exclusively implemented in industrial mixers. Despite its
omnipresence in applications, alternative strategies are often equally
or more effective, foremost among them vortex shedding due to unsteady
and abrupt motion of the stirrers, affably denoted as the {\bf{vortex
    cannon}} strategy (figures~\ref{fig:Strategies}{\it{d-f}}). In
this case, the stirrer generates a sequence of startup and stopping
vortices by rapid oscillations or abrupt directional changes along the
circular paths. The shed vortices then act as effective autonomous
mixers that, once they reach the initial or distorted interface,
further deform the passive scalar field and locally (and globally)
reduce the mix-norm. In this manner, a single stirrer can clone 'fluid
stirrers' (shed vortices) and thus multiply its mixing
effectiveness. In a further possible strategy, vortices can be
generated in the fluid that collide with each other and thus generate
filaments, which are then subjected to more rapid diffusion and
homogenisation (figures~\ref{fig:Strategies}{\it{g-i}}). Of course,
this {\bf{vortex collision}} strategy is strongly dependent on the
initial condition of the passive scalar -- and for this reason, less
transferable to a general, realistic mixing strategy --, nonetheless,
within our computational framework, it is a viable and pervasive
strategy utilised by our direct-adjoint algorithm. A far more
transferable mixing strategy is the collision of vortical structures
with the outer wall (figures~\ref{fig:Strategies}{\it{j-l}}) whereby a
large fluid element is broken up into smaller elements which further
interact with other vortices and are subject to increased diffusion
due to the breakdown in scales. Finally, the embedded physical
stirrers can themselves interplay with the vortical structures they
generate, acting as {\bf{obstructions}} in the path of vortices
(figures~\ref{fig:Strategies}{\it{m-o}}). A collision between a
stirrer and a vortex will split the vortex and yield smaller scales,
hence contributing to a decrease in the mix-norm. This final strategy
will continue to cause a moderate breakdown in scales, even after the
control window has closed and no more stirring motion is allowed.

Given these five fundamental strategies, illustrated in
figure~\ref{fig:Strategies} with samples from our simulations, the
direct-adjoint looping algorithm will select from and combine these
options into a coherent strategy, given the chosen parameters and
user-specified constraints.

\subsection{ Overview of test cases}

We will consider six cases, grouped into three examples. Each example
consists of a short-time strategy with a rather limited control
horizon of $T_{\rm{control}}=1$ and a long-time strategy with a more
generous horizon of $T_{\rm{control}}=8.$ These two
$T_{\rm{control}}$-settings will impose noticeable constraints on the
choice of strategies, the interplay of dynamic processes and the
feasibility of the final protocol. The three examples further
distinguish themselves by the number of external constraints: starting
with pure energy constraints, via energy and velocity constraints, to
energy, velocity and acceleration constraints. Along this course of
action, algorithmic requisites and physical requirements will be
encountered and discussed.

Convergence of the iterative scheme is principally governed by
constraints imposed on the optimisation problem. First, the nonlinear
nature of the governing equations precludes a guarantee to converge
towards a global minimum; only a local minimum can be expected.
More importantly, additional constraints on the stirrers, such as
energy, velocity or acceleration bounds, can convert a semi-norm to
a full-norm optimisation problem. Semi-norm optimisation
problems~\citep{Foures2012,Blumenthal2017} are 'open-ended' in the
sense that the stirrer velocities increase without bounds, while the
mixing measure steadily improves. In this case, the iterative optimisation
scheme terminates when the adjoint variables, due to excessive direct
velocities, no longer furnish useful gradient information for further
improvement. In other words, the optimisation comes to an end when the
signal-to-noise ratio for the gradient information drops to a value
near unity, even before the cost-functional gradient attains a
small value. In the full-norm case, convergence is achieved when the
cost-functional gradient falls below a small, user-specified threshold,
indicating that a {\it{local}} minimum has been reached. In brief,
iterations are halted when either the cost-functional gradient falls
below a small threshold or the signal-to-noise ratio of the adjoint
gradient information reaches unity -- whichever scenario comes first.


Since mixing enhancement based on complex stirring strategies is a
highly dynamic process -- based on a rapid sequence of abundant
vortical features --, a set of static snapshots cannot do justice to
the intricacies of an optimised mixing protocol. For this reason, we
urge the reader to turn to the animations in the supplemental
material.








\subsection{Cases 1 and 2: optimisation under energy constraints}

The first two cases follow a common procedure whereby the cost
functional (mix-norm of the passive scalar) of the constrained
optimisation is minimised, subject to a penalisation of the control
energy that accomplishes this minimum. Since the stirrers' kinetic
energy is a measure of effort that goes into the mixing process, we
add a corresponding term to the pure mix-norm cost functional. As a
consequence, the energy expended by the stirrers is bounded to a
user-specified value.

\begin{figure}
  \centering
  \begin{tabular}{ccc}
    \begin{tikzpicture}[>=stealth]
      \draw (0,0) node[inner sep=0] {\includegraphics[height=0.27\textwidth]{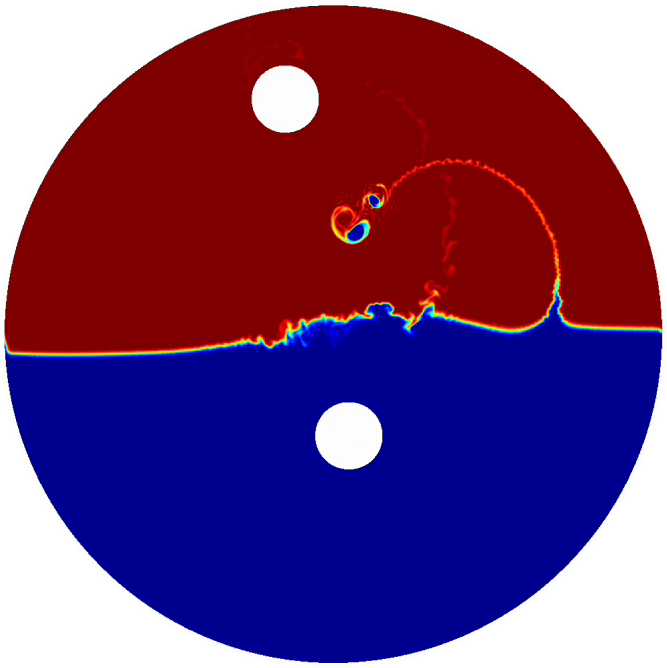}};
      \draw [gray!50,thin,domain=90:105,samples=100,->] plot ({1.3*cos(\x)}, {1.3*sin(\x)});
      \draw [gray!50,thin,domain=-90:-75,samples=100,->] plot ({0.575*cos(\x)}, {0.575*sin(\x)});

      \draw [white,dashed,thick,domain=120:150,samples=100,->] plot ({1.45*cos(\x)}, {1.45*sin(\x)});
      \draw [white,dashed,thick,domain=120:150,samples=100,->] plot ({1.2*cos(\x)}, {1.2*sin(\x)});
      \draw [white,dashed,thick,domain=-60:-25,samples=100,->] plot ({0.7*cos(\x)}, {0.7*sin(\x)});
      \draw [white,dashed,thick,domain=-60:-25,samples=100,->] plot ({0.4*cos(\x)}, {0.4*sin(\x)});

      \draw[black] (-1.5,-1.5) node {$(a)$};
      \draw[anchor=east] (2,-1.65) node {\scriptsize{$t=0.55$}};
    \end{tikzpicture} & \hspace{0truecm}
    \begin{tikzpicture}[>=stealth]
      \draw (0,0) node[inner sep=0] {\includegraphics[height=0.27\textwidth]{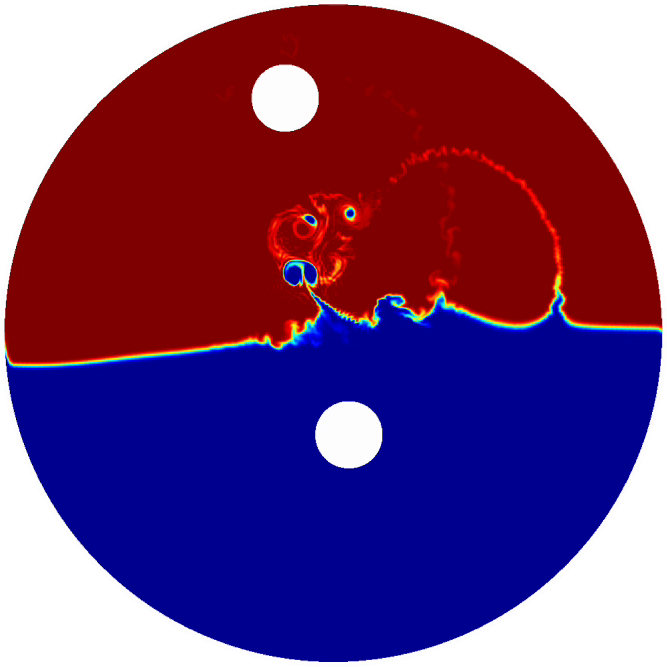}};

      \draw[white,thick,densely dashed,->] (0.15,0.9) to [out=205,in=90] (-0.15,0.45);
      \draw[white,thick,densely dashed,->] (0.15,0) to [out=145,in=-90] (-0.15,0.45);

      \draw[black] (-1.5,-1.5) node {$(b)$};
      \draw[anchor=east] (2,-1.65) node {\scriptsize{$t=0.67$}};
    \end{tikzpicture} & \hspace{0truecm}
    \begin{tikzpicture}[>=stealth]
      \draw (0,0) node[inner sep=0] {\includegraphics[height=0.27\textwidth]{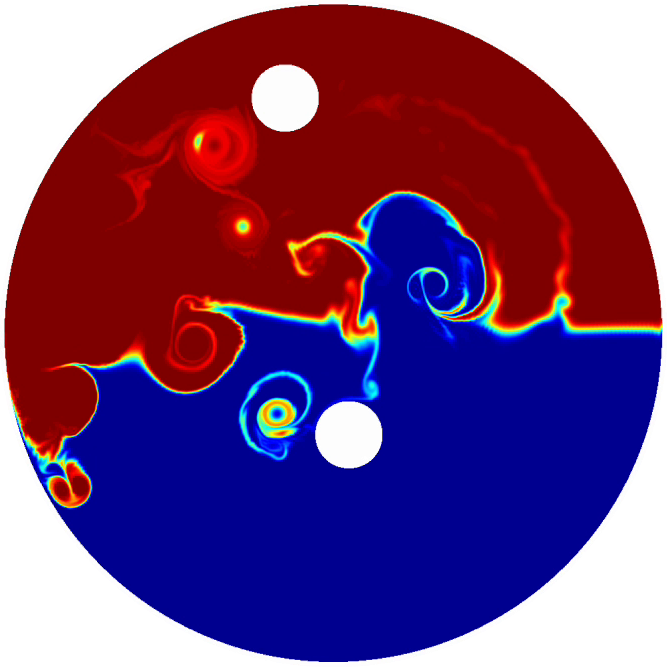}};

      \draw[white,thick,densely dashed,->] (-1.3,-1) to [out=-45,in=180] (-0.7,-1.2);
      \draw[white,thick,densely dashed,->] (-0.9,-0.4) to [out=-60,in=180] (-0.5,-0.6);

      \draw[black] (-1.5,-1.5) node {$(c)$};
      \draw[anchor=east] (2,-1.65) node {\scriptsize{$t=1.48$}};
    \end{tikzpicture} \\
    \begin{tikzpicture}[>=stealth]
      \draw (0,0) node[inner sep=0] {\includegraphics[height=0.27\textwidth]{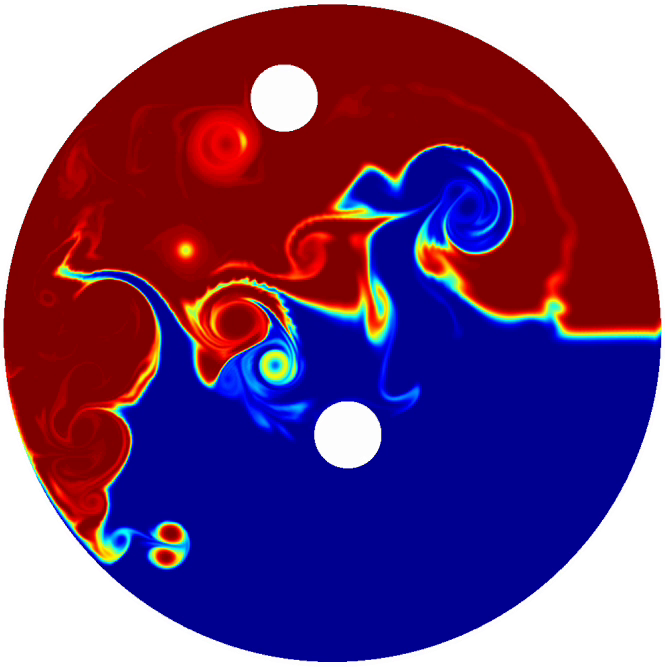}};

      \draw[white,thick,densely dashed,->] (-0.7,-1.2) to [out=0,in=-135] (-0.1,-0.9);
      \draw[white,thick,densely dashed,->] (-0.3,0) to [out=35,in=-60] (-0.1,0.9);

      \draw[black] (-1.5,-1.5) node {$(d)$};
      \draw[anchor=east] (2,-1.65) node {\scriptsize{$t=2.54$}};
    \end{tikzpicture} & \hspace{0truecm}
    \begin{tikzpicture}[>=stealth]
      \draw (0,0) node[inner sep=0] {\includegraphics[height=0.27\textwidth]{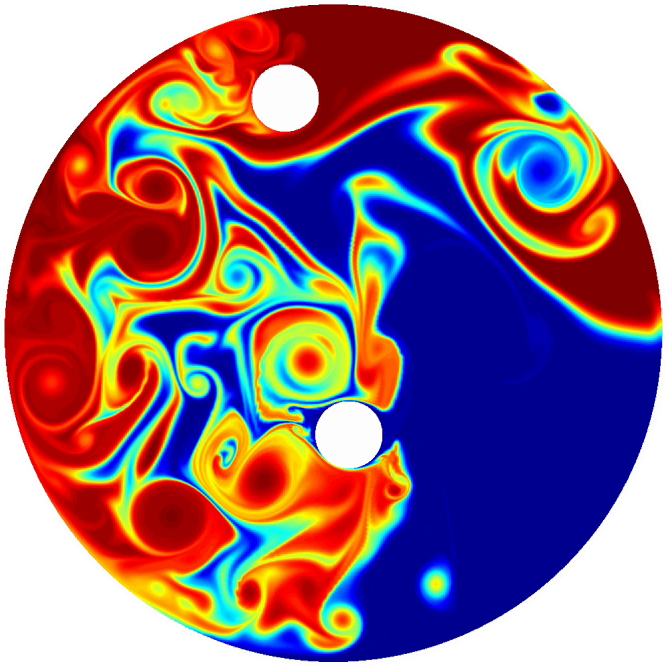}};

      \draw[white,thick,densely dashed,->] (-0.5,-0.7) to [out=45,in=-110] (-1.5,0.6);
      \draw[white,thick,densely dashed,->] (0.1,-0.9) to [out=-5,in=45] (0.9,-1.4);
      \draw[white,thick,densely dashed,->] (0.1,-0.8) to [out=-10,in=-120] (1.4,-0.3);

      \draw[black] (-1.5,-1.5) node {$(e)$};
      \draw[anchor=east] (2,-1.65) node {\scriptsize{$t=8$}};
    \end{tikzpicture} & \hspace{0truecm}
    \begin{tikzpicture}
      \draw (0,0) node[inner sep=0] {\includegraphics[height=0.27\textwidth]{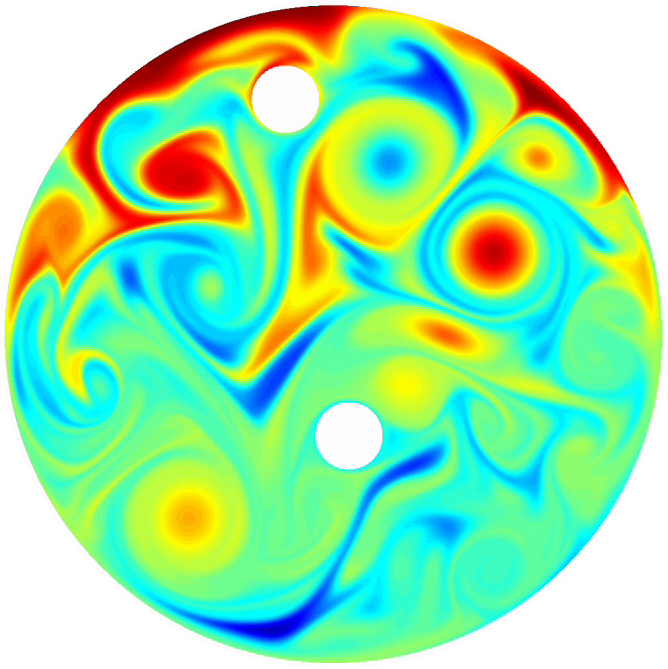}};
      \draw[black] (-1.5,-1.5) node {$(f)$};
      \draw[anchor=east] (2,-1.65) node {\scriptsize{$t=32$}};
    \end{tikzpicture}
  \end{tabular}
  \caption{\label{fig:EnConstraint1} Mixing optimisation based on only
    energy constraints for the stirrers. The time horizon for applying
    control is $T_{\rm{control}}=1.$ Shown are iso-contours of the
    passive scalar at selected instances. The optimisation algorithm
    includes information over a time window of $T_{\rm{info}}=8.$}
\end{figure}
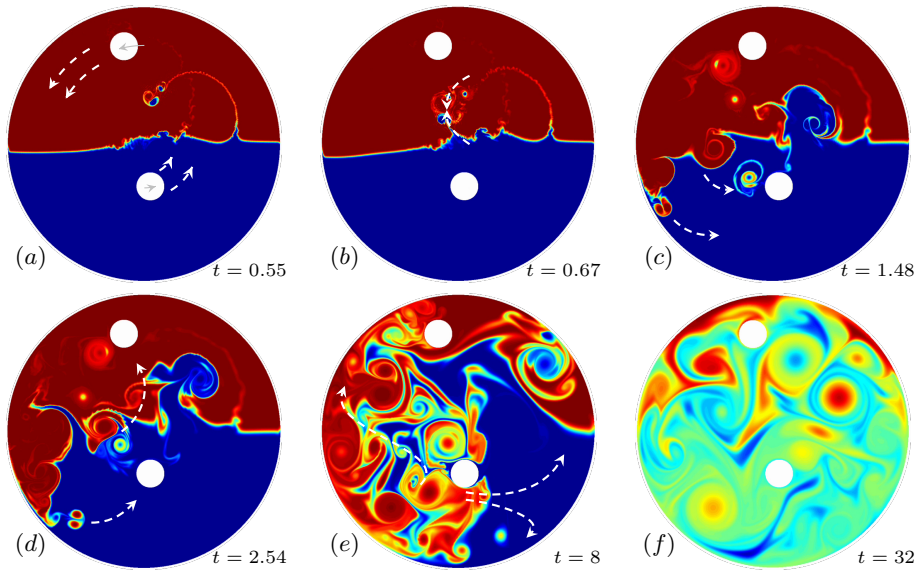

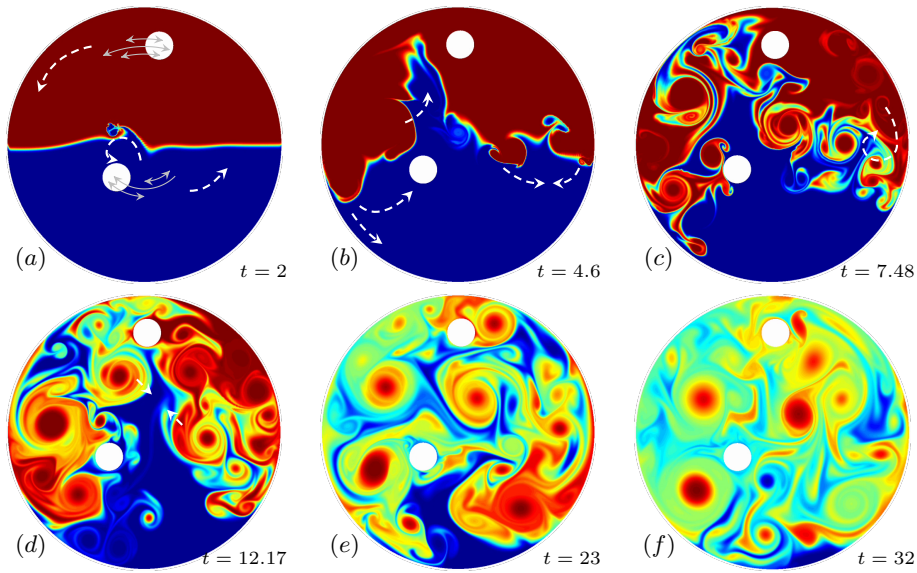
\begin{figure}
  \centering
  \begin{tabular}{ccc}
    \begin{tikzpicture}[>=stealth]
      \draw (0,0) node[inner sep=0] {\includegraphics[height=0.27\textwidth]{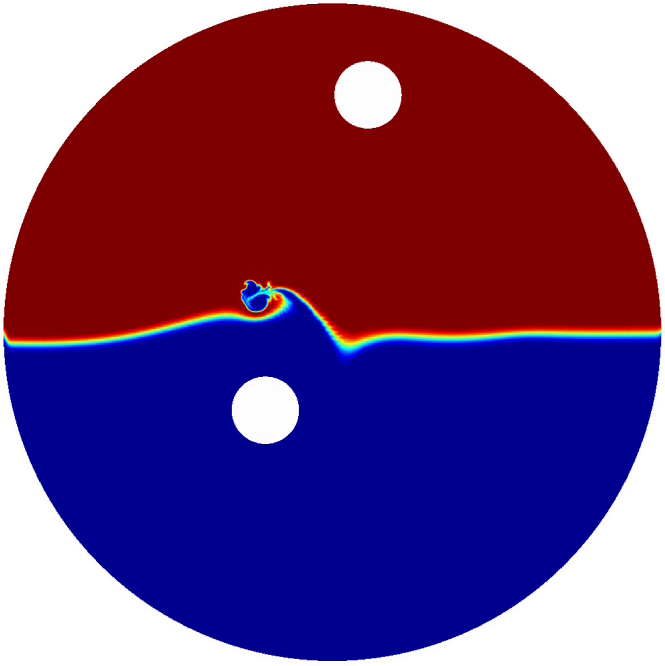}};
      \draw [gray!50,thin,domain=80:100,samples=100,<->] plot ({1.4*cos(\x)}, {1.4*sin(\x)});
      \draw [gray!50,thin,domain=75:115,samples=100,<->] plot ({1.3*cos(\x)}, {1.3*sin(\x)});
      \draw [gray!50,thin,domain=80:105,samples=100,<->] plot ({1.2*cos(\x)}, {1.2*sin(\x)});
      \draw [gray!50,thin,domain=-135:-45,samples=100,<-] plot ({0.575*cos(\x)}, {0.575*sin(\x)});
      \draw [gray!50,thin,domain=-135:-90,samples=100,<->] plot ({0.675*cos(\x)}, {0.675*sin(\x)});
      \draw [gray!50,thin,domain=-90:-45,samples=100,<->] plot ({0.475*cos(\x)}, {0.475*sin(\x)});

      \draw[white,thick,densely dashed,->] (-0.05,-0.2) to [out=100,in=0] (-0.3,0.1) to [out=180,in=90] (-0.5,-0.1) to [out=270,in=170] (-0.35,-0.2);
      \draw[white,thick,densely dashed,->] (0.6,-0.6) to [out=0,in=-135] (1.1,-0.3);
      \draw[white,thick,densely dashed,->] (-0.7,1.3) to [out=190,in=70] (-1.4,0.7);

      \draw[black] (-1.5,-1.5) node {$(a)$};
      \draw[anchor=east] (2,-1.65) node {\scriptsize{$t=2$}};
    \end{tikzpicture} & \hspace{0truecm}
    \begin{tikzpicture}[>=stealth]
      \draw (0,0) node[inner sep=0] {\includegraphics[height=0.27\textwidth]{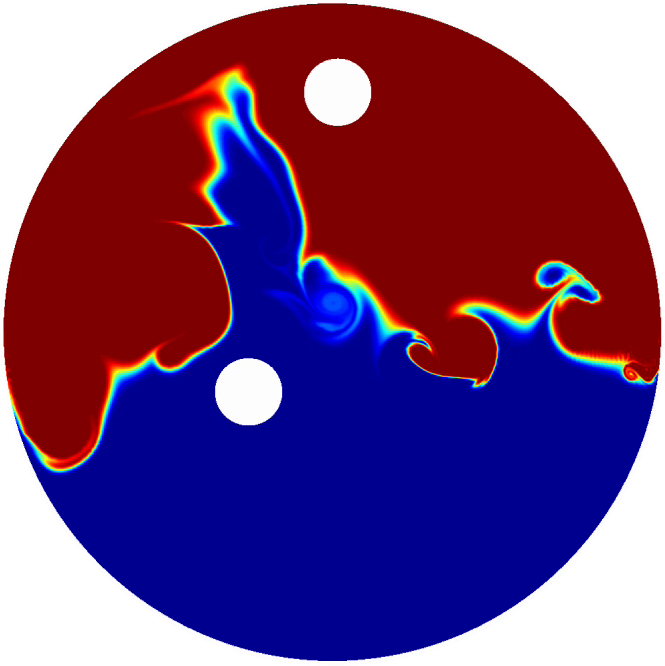}};

      \draw[white,thick,densely dashed,->] (0.6,-0.3) to [out=-40,in=180] (1.15,-0.5);
      \draw[white,thick,densely dashed,->] (1.6,-0.3) to [out=-110,in=0] (1.25,-0.5);
      \draw[white,thick,densely dashed,->] (-1.3,-0.8) to [out=-40,in=-135] (-0.6,-0.6);
      \draw[white,thick,densely dashed,->] (-1.4,-0.9) to [out=-55,in=145] (-1,-1.3);
      \draw[white,thick,densely dashed,->] (-0.7,0.3) to [out=20,in=-90] (-0.4,0.7);

      \draw[black] (-1.5,-1.5) node {$(b)$};
      \draw[anchor=east] (2,-1.65) node {\scriptsize{$t=4.6$}};
    \end{tikzpicture} & \hspace{0truecm}
    \begin{tikzpicture}[>=stealth]
      \draw (0,0) node[inner sep=0] {\includegraphics[height=0.27\textwidth]{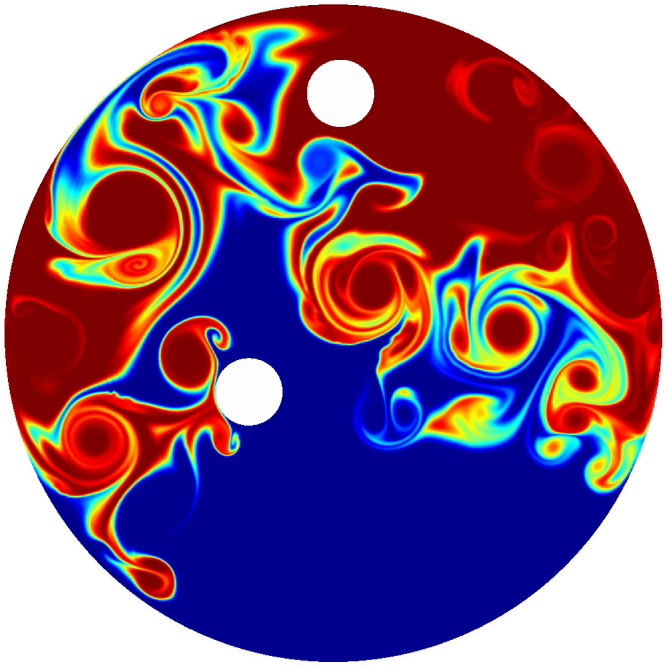}};

      \draw[white,thick,densely dashed,->] (1.5,0.5) to [out=-55,in=90] (1.65,0) to [out=-90,in=0] (1.4,-0.2) to [out=180,in=-90] (1.2,0) to [out=90,in=-135] (1.4,0.2);

      \draw[black] (-1.5,-1.5) node {$(c)$};
      \draw[anchor=east] (2,-1.65) node {\scriptsize{$t=7.48$}};
    \end{tikzpicture} \\
    \begin{tikzpicture}[>=stealth]
      \draw (0,0) node[inner sep=0] {\includegraphics[height=0.27\textwidth]{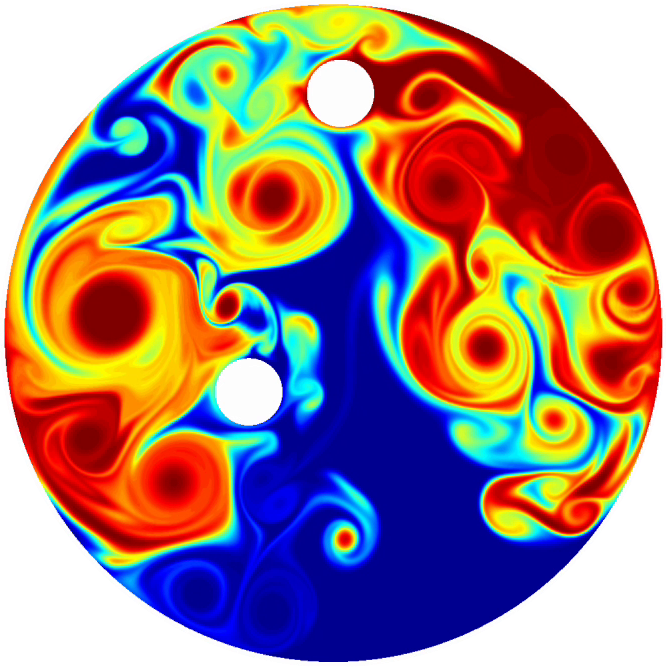}};

      \draw[white,thick,densely dashed,->] (0.5,0.1) -- (0.3,0.3);
      \draw[white,thick,densely dashed,->] (-0.1,0.7) -- (0.1,0.5);

      \draw[black] (-1.5,-1.5) node {$(d)$};
      \draw[anchor=east] (2,-1.65) node {\scriptsize{$t=12.17$}};
    \end{tikzpicture} & \hspace{0truecm}
    \begin{tikzpicture}
      \draw (0,0) node[inner sep=0] {\includegraphics[height=0.27\textwidth]{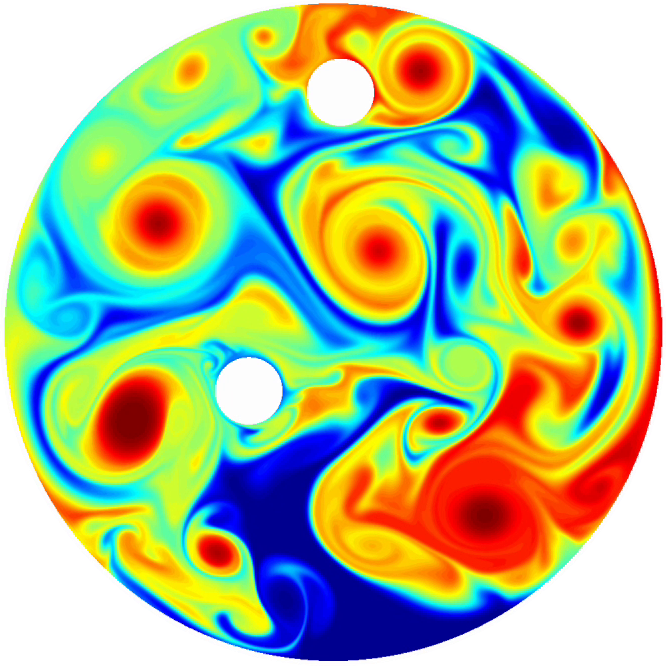}};
      \draw[black] (-1.5,-1.5) node {$(e)$};
      \draw[anchor=east] (2,-1.65) node {\scriptsize{$t=23$}};
    \end{tikzpicture} & \hspace{0truecm}
    \begin{tikzpicture}
      \draw (0,0) node[inner sep=0] {\includegraphics[height=0.27\textwidth]{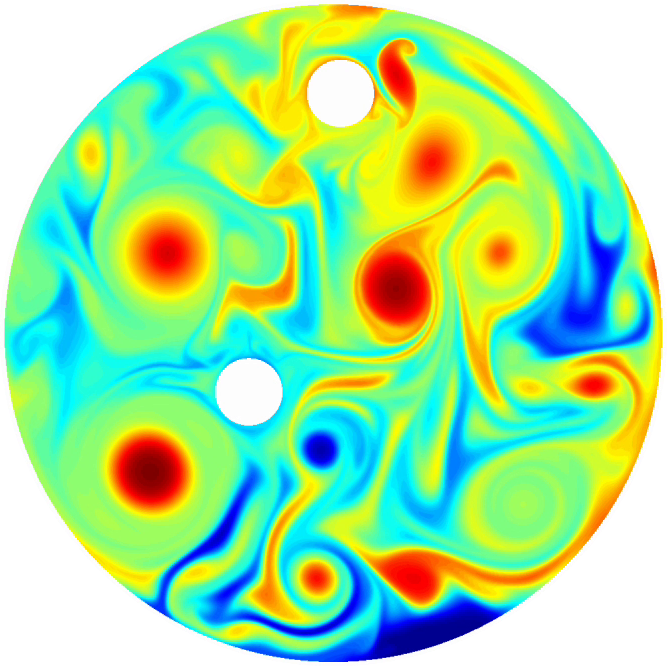}};
      \draw[black] (-1.5,-1.5) node {$(f)$};
      \draw[anchor=east] (2,-1.65) node {\scriptsize{$t=32$}};
    \end{tikzpicture}
  \end{tabular}
  \caption{\label{fig:EnConstraint8} Same as
    figure~\ref{fig:EnConstraint1}, but with an extended control
    window of $T_{\rm{control}}=8.$}
\end{figure}

Figure~\ref{fig:EnConstraint1} displays the results of our
optimisation, visualised by iso-contours of the passive scalar
$\theta$ at selected time instances. The control horizon is
$T_{\rm{control}}=1.$ We observed that the optimisation does not
utilise the `plunging' option, as the stirrers remain nearly at their
initial position. Instead, the entire energy available to the stirrers
is used up in a rapid start-and-stop motion which initially causes
multiple small-scale shed vortices that distort the plane interface,
collide into each other and the stirrers, and merge into larger-scale
vortex structures which eventually achieve good mixing. It is
important to stress that for the calculation of this short-time mixing
strategy, information about the full dynamics up to $T_{\rm{info}}=8$
has been incorporated into the optimisation. In other words, the
consequences of the limited stirring protocol up to $T_{\rm{info}}=8$
are known to the optimisation, and adjustments to the control strategy
can be made that affect the vortex dynamics beyond its active control
window. The evolution of the passive scalar between $T_{\rm{info}}=8$
and $T_{\rm{sim}}=32,$ however, is neither designed nor recognized by
the optimzation algorithm; it simply plays out according to the action
taken during the control and optimisation windows. We include this
further evolution to underline our choice of the mix-norm as the
mixedness measure, whose optimisation produces the small scales that
are subsequently diffused during this `cool-off' window.

We conclude that the absence of any plunging option points at the
suboptimality of this particular strategy in achieving an enhanced mixing
process. It is thus not pursued as a viable option by the
direct-adjoint optimisation technique. In the interpretation of these
results, it may be tempting to conclude that a different initial
placement of the cylinders -- closer to the initial interface -- would
have resulted in strategies that included plunging. However, a
simulation of the same case (not shown), with the two cylinders
starting immersed in the initial interface, came to the same
conclusion: while, by design, there is a small amount of plunging in
this case, the vast majority of the mix-norm reduction has been
accomplished by the shedding of start-and-stop vortices by a vigorous
oscillatory motion of either cylinder and a subsequent collision of
the generated vortices. The utilization of the stirrers' energy to
shed small ``vortical stirrers'' is a better strategy than the
distortion of the interface by simply moving through it with the
stirrers.

The later part of the stirring strategy includes vortex collisions
(see figure~\ref{fig:EnConstraint1}{\it{b}}), obstruction by the stirrers
(see figures~\ref{fig:EnConstraint1}{\it{c,d}}) and collision with the
outer wall (see figure~\ref{fig:EnConstraint1}{\it{e}}) to yield a
well-mixed state at the end of the simulation horizon
(figure~\ref{fig:EnConstraint1}{\it{f}}).

Increasing the control horizon from $T_{\rm{control}}=1$ to
$T_{\rm{control}}=8$ leads to similar conclusions, even though the
stirring action by the cylinders is less abrupt and jarring. Still,
the bulk of the mixing action is achieved by shedding start-and-stop
vortices which collide with themselves, secondary vortices and the
wall to produce a mixed state in the end. Again, the absence of
plunging is noteworthy. This is even more remarkable, as the increased
control time horizon would certainly allow the stirrers to approach
and reach the interface; yet, they remain close to their initial
position.

In both cases, the strategy found by the direct-adjoint optimisation
technique will yield increasingly larger velocities, as long as the
integrated energy is constant. Eventually, the energy expenditure
becomes more and more localized in time, with the stirrers barely
moving. This optimisation route is a logical consequence of our
current setup. It is closely connected to the semi-norm problem (see
\cite{Foures2012,Blumenthal2017}): the mix-norm only contains the
passive scalar $\theta$, but does not account for the other dynamic
variable, the velocities, in the optimisation. The energy of the
stirrers is not sufficient to arrive at realistic stirring protocols
that could be implemented in an experimental or industrial setting. To
ensure applicability of our stirrer strategies to real-life settings,
additional constraints are required.

\subsection{Cases 3 and 4: optimisation under energy and velocity constraints}

Following the findings of the previous section, in the next examples
we limit the velocity of the stirrers to avoid excessive values of
${\bf{u}}_{s,i}.$ This capping of the velocity is implemented by a
projection of the raw cost-functional gradient onto control strategies
that satisfy the given constraints (for details of this projection
technique see~\cite{Foures2014}). The resulting limit on the stirrer
velocities provides a longer time window (up to the control horizon
$T_{\rm{control}}$) over which the specified energy can be
expended. As a consequence, an extended and smoother movement of the
stirrers is expected.

\begin{figure}
  \centering
  \begin{tabular}{ccc}
    \begin{tikzpicture}[>=stealth]
      \draw (0,0) node[inner sep=0] {\includegraphics[height=0.27\textwidth]{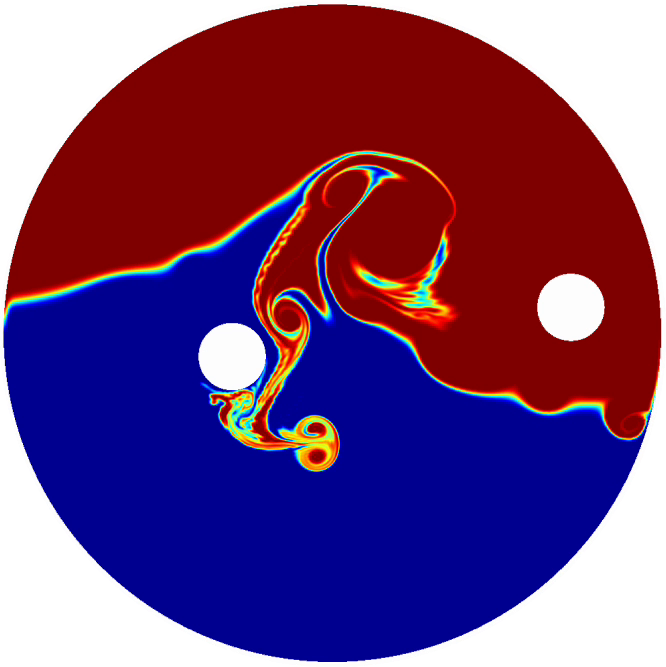}};

      \draw [gray!50,thin,domain=80:100,samples=100,<->] plot ({1.4*cos(\x)}, {1.4*sin(\x)});
      \draw [gray!50,thin,domain=90:5,samples=100,->] plot ({1.3*cos(\x)}, {1.3*sin(\x)});
      \draw [gray!50,thin,domain=-90:-165,samples=100,->] plot ({0.575*cos(\x)}, {0.575*sin(\x)});

      \draw[white,thick,densely dashed,->] (0.1,-0.6) to [out=0,in=-90] (0.3,-0.3) to [out=90,in=-15] (0,0);
      \draw[white,thick,densely dashed,->] (0.4,0.6) to [out=0,in=90] (0.8,0.4) to [out=-90,in=0] (0.5,0.2);

      \draw[black] (-1.5,-1.5) node {$(a)$};
      \draw[anchor=east] (2,-1.65) node {\scriptsize{$t=1.58$}};
    \end{tikzpicture} & \hspace{0truecm}
    \begin{tikzpicture}[>=stealth]
      \draw (0,0) node[inner sep=0] {\includegraphics[height=0.27\textwidth]{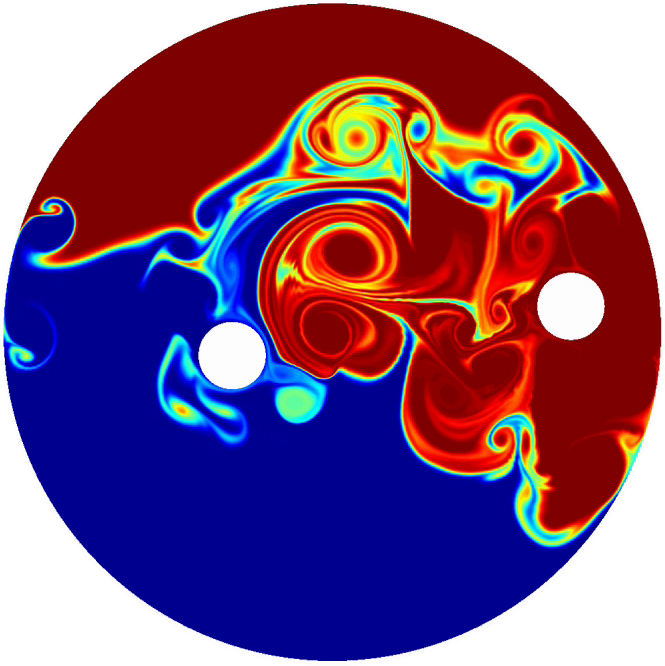}};

      \draw[white,thick,densely dashed,->] (0.6,-0.6) to [out=-135,in=0] (0,-0.8);
      \draw[white,thick,densely dashed,->] (-0.1,0.3) to [out=160,in=-90] (-0.5,0.9) to [out=90,in=160] (0.5,1.45);

      \draw[black] (-1.5,-1.5) node {$(b)$};
      \draw[anchor=east] (2,-1.65) node {\scriptsize{$t=3.64$}};
    \end{tikzpicture} & \hspace{0truecm}
    \begin{tikzpicture}[>=stealth]
      \draw (0,0) node[inner sep=0] {\includegraphics[height=0.27\textwidth]{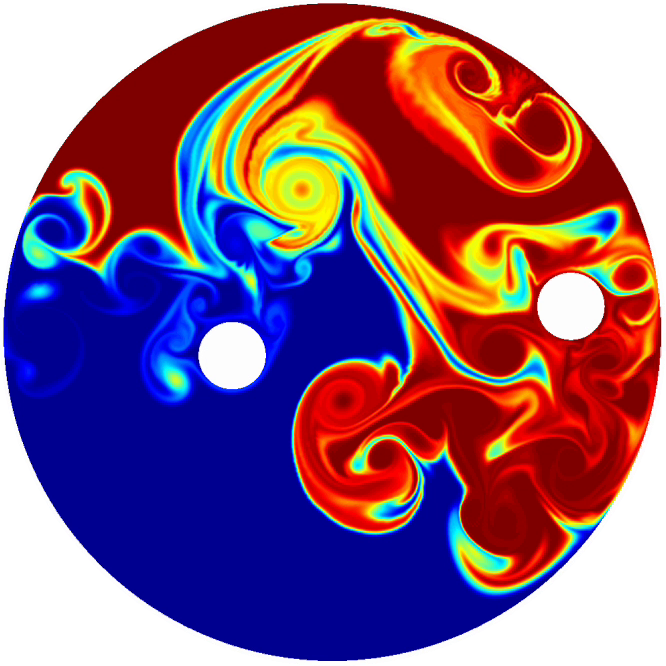}};

      \draw[white,thick,densely dashed,->] (-0.3,-0.5) to [out=155,in=90] (-1.1,-0.8) to [out=-90,in=180] (-0.6,-1.1);
      \draw[white,thick,densely dashed,->] (1.2,0.9) to [out=-45,in=90] (1.3,0.6) to [out=-90,in=0] (1,0.4) to [out=180,in=-90] (0.7,0.7) to [out=90,in=-135] (0.9,1);

      \draw[black] (-1.5,-1.5) node {$(c)$};
      \draw[anchor=east] (2,-1.65) node {\scriptsize{$t=5.18$}};
    \end{tikzpicture} \\
    \begin{tikzpicture}[>=stealth]
      \draw (0,0) node[inner sep=0] {\includegraphics[height=0.27\textwidth]{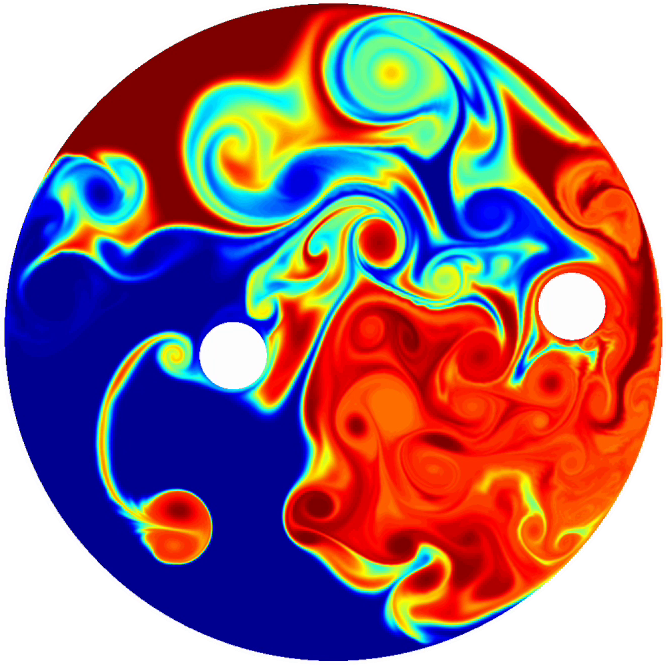}};

      \draw[white,thick,densely dashed,->] (0,-0.4) to [out=135,in=-90] (-0.3,0.3);
      \draw[white,thick,densely dashed,->] (-0.6,-1) -- (-0.45,-0.9);
      \draw[white,thick,densely dashed,->] (-0.3,-1) -- (-0.45,-0.9);
      \draw[white,thick,densely dashed,->] (-0.45,-0.85) to [out=95,in=0] (-0.8,-0.4) to [out=180,in=75] (-1.05,-0.7);

      \draw[black] (-1.5,-1.5) node {$(d)$};
      \draw[anchor=east] (2,-1.65) node {\scriptsize{$t=8.51$}};
    \end{tikzpicture} & \hspace{0truecm}
    \begin{tikzpicture}
      \draw (0,0) node[inner sep=0] {\includegraphics[height=0.27\textwidth]{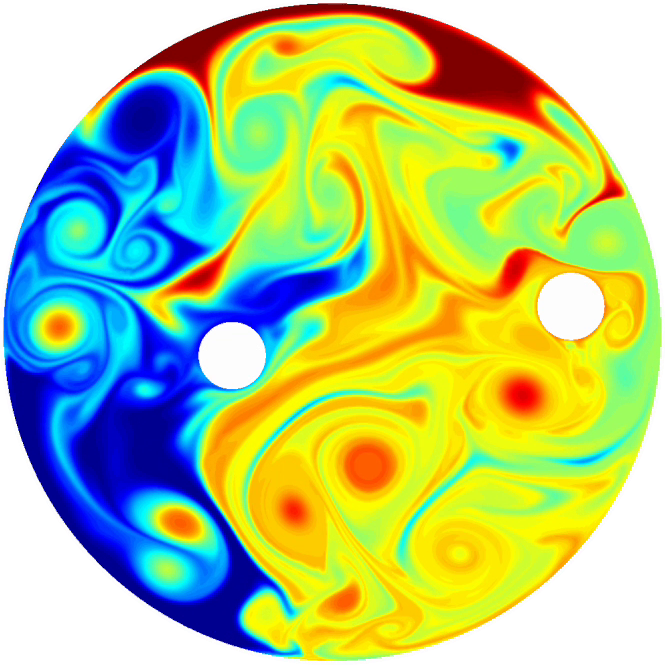}};
      \draw[black] (-1.5,-1.5) node {$(e)$};
      \draw[anchor=east] (2,-1.65) node {\scriptsize{$t=18.78$}};
    \end{tikzpicture} & \hspace{0truecm}
    \begin{tikzpicture}
      \draw (0,0) node[inner sep=0] {\includegraphics[height=0.27\textwidth]{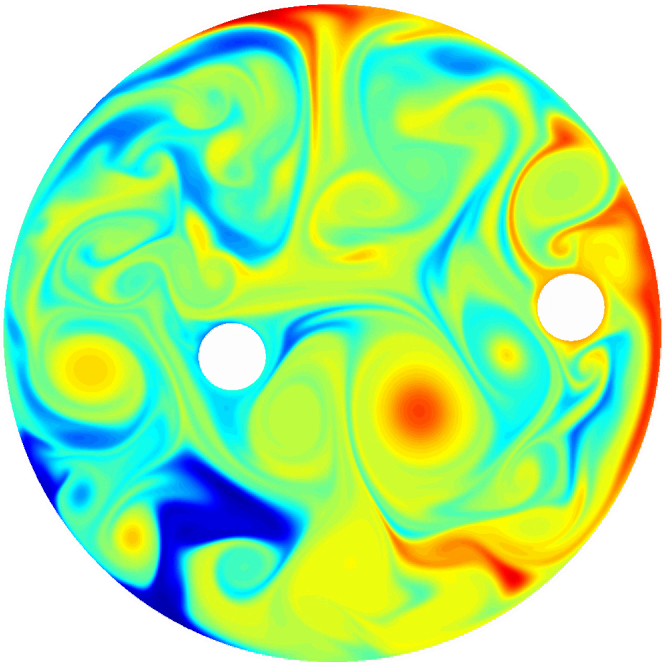}};
      \draw[black] (-1.5,-1.5) node {$(f)$};
      \draw[anchor=east] (2,-1.65) node {\scriptsize{$t=32$}};
    \end{tikzpicture}
  \end{tabular}
  \caption{\label{fig:VelConstraint1} Mixing optimisation based on
    energy and velocity constraints for the stirrers. The time horizon
    for applying control is $T_{\rm{control}}=1.$ Shown are
    iso-contours of the passive scalar at selected instances. The
    optimisation algorithm includes information over a time window of
    $T_{\rm{info}}=8.$}
\end{figure}
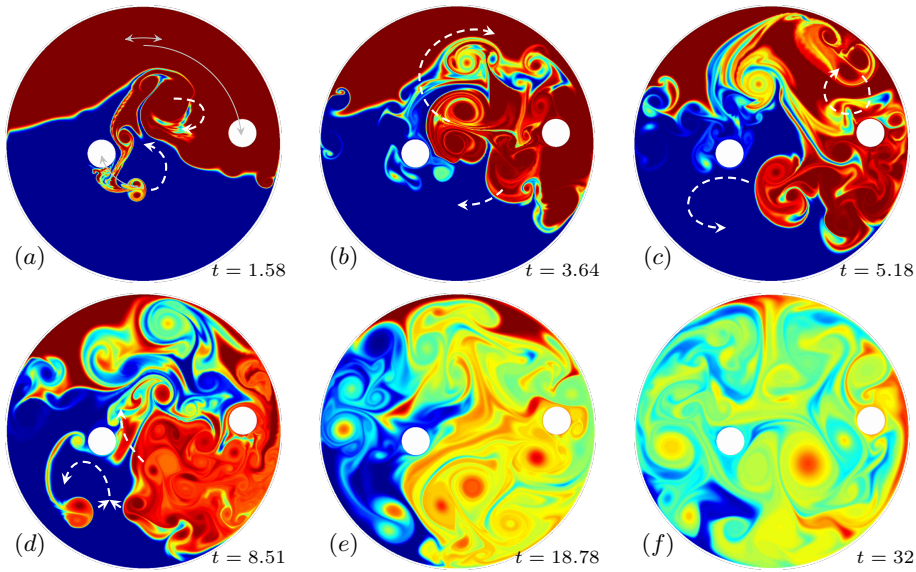

For the shorter control horizon $T_{\rm{control}}=1,$
figure~\ref{fig:VelConstraint1} shows the outcome of our
optimisation. The top stirrer starts by an oscillatory motion,
creating start-and-stop vortices. The capping of its velocity,
however, keeps the shed vortices within bounds; nonetheless, the
optimality of the ``vortex cannon'' strategy can still be exploited.
Both stirrers then move closer to the (already distorted)
interface. But rather than plunging through it, they abruptly stop
short of it and let the overtaking stop-vortices carry out the
distortion of the interface and the subsequent mixing. Again, the
optimisation algorithm selects the mixing by shed vortices over the
plunging of the stirrers through the interface. The remaining mixing
process is characterized by vortex collision (see
figures~\ref{fig:VelConstraint1}{\it{a,d}}), collision with the wall (see
figure~\ref{fig:VelConstraint1}{\it{c}}) and stirrer obstruction (see
figures~\ref{fig:VelConstraint1}{\it{c,d}}).

\begin{figure}
  \centering
  \begin{tabular}{ccc}
    \begin{tikzpicture}[>=stealth]
      \draw (0,0) node[inner sep=0] {\includegraphics[height=0.27\textwidth]{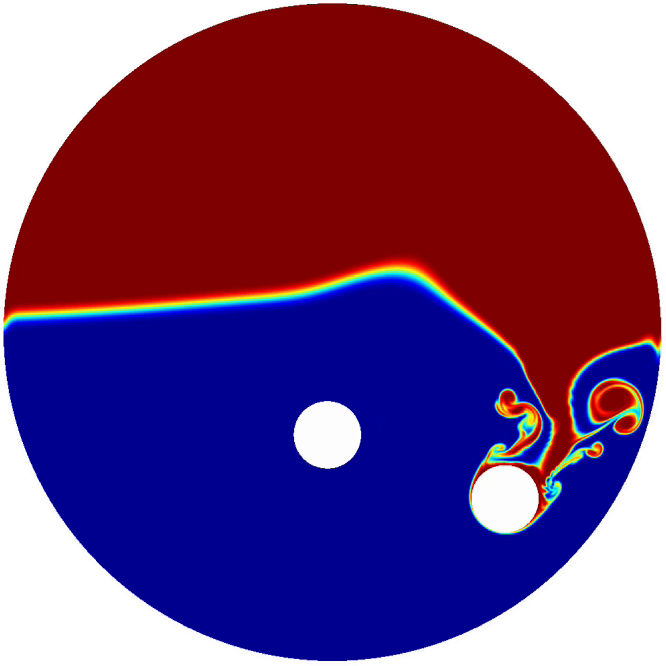}};

     \draw [gray!50,thin,domain=90:5,samples=100,->] plot ({1.3*cos(\x)}, {1.3*sin(\x)});
     \draw [gray!50,thin,domain=0:-20,samples=100,->] plot ({1.3*cos(\x)}, {1.3*sin(\x)});
     \draw [gray!50,thin,domain=-25:-75,samples=100,->] plot ({1.3*cos(\x)}, {1.3*sin(\x)});
     \draw [gray!50,thin,domain=-65:-85,samples=100,<->] plot ({1.4*cos(\x)}, {1.4*sin(\x)});

      \draw [gray!50,thin,domain=-90:-55,samples=100,->] plot ({0.475*cos(\x)}, {0.475*sin(\x)});
      \draw [gray!50,thin,domain=-55:-125,samples=100,<->] plot ({0.575*cos(\x)}, {0.575*sin(\x)});
      \draw [gray!50,thin,domain=-125:-90,samples=100,->] plot ({0.675*cos(\x)}, {0.675*sin(\x)});

      \draw[white,thick,densely dashed,->] (1.7,-0.2) to [out=95,in=0] (1.3,0.1) to [out=180,in=90] (0.9,-0.3) to [out=-90,in=135] (1.1,-0.6);

      \draw[black] (-1.5,-1.5) node {$(a)$};
      \draw[anchor=east] (2,-1.65) node {\scriptsize{$t=3.38$}};
    \end{tikzpicture} & \hspace{0truecm}
    \begin{tikzpicture}[>=stealth]
      \draw (0,0) node[inner sep=0] {\includegraphics[height=0.27\textwidth]{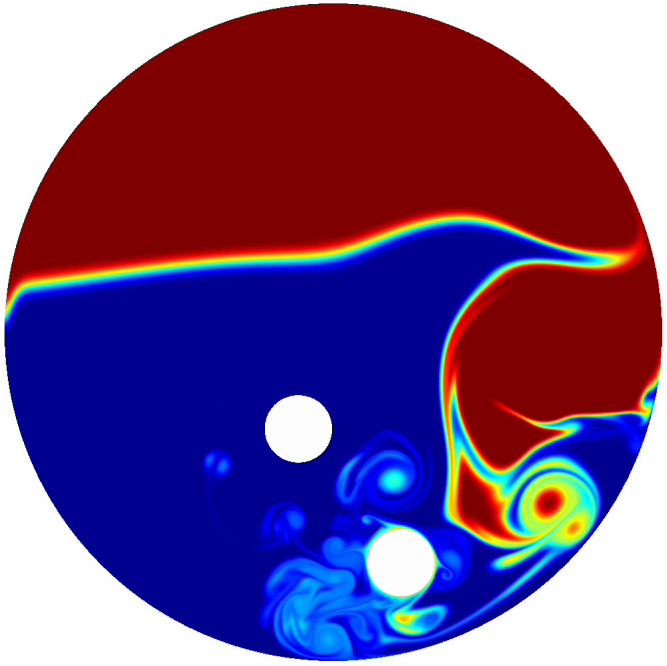}};

      \draw[white,thick,densely dashed,->] (1.5,0.7) to [out=-65,in=90] (1.6,0.2) to [out=-90,in=25] (0.4,-0.8);
      \draw[white,thick,densely dashed,->] (1.1,-0.2) to [out=-135,in=0] (0.6,-0.5) to [out=180,in=-105] (0.3,-0.1);
      \draw[white,thick,densely dashed,->] (-0.5,-1.5) to [out=165,in=-45] (-0.9,-1.3);

      \draw[black] (-1.5,-1.5) node {$(b)$};
      \draw[anchor=east] (2,-1.65) node {\scriptsize{$t=6.85$}};
    \end{tikzpicture} & \hspace{0truecm}
    \begin{tikzpicture}[>=stealth]
      \draw (0,0) node[inner sep=0] {\includegraphics[height=0.27\textwidth]{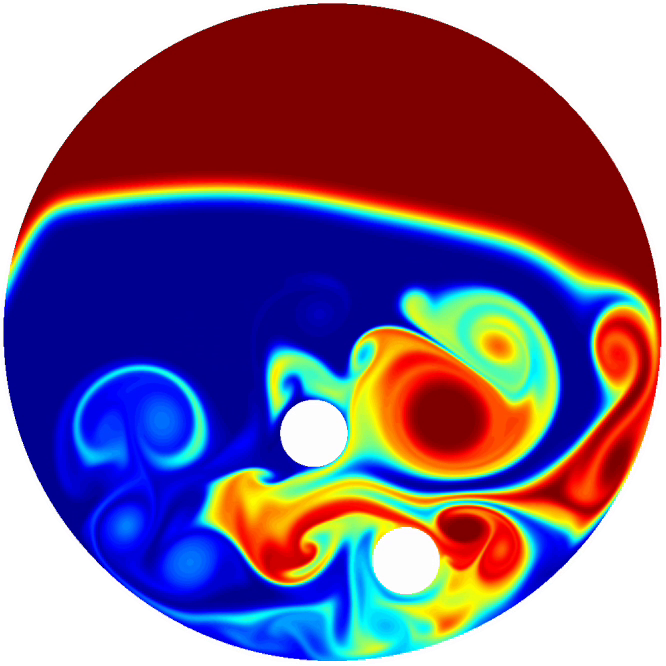}};

      \draw[white,thick,densely dashed,->] (-0.5,-1.2) to [out=-155,in=90] (-0.6,-1.35) to [out=-90,in=180] (-0.2,-1.6);
      \draw[white,thick,densely dashed,->] (-0.7,-1.1) to [out=180,in=-90] (-1.1,-0.6) to [out=90,in=-155] (-0.7,-0.1);
      \draw[white,thick,densely dashed,->] (0.9,-0.3) to [out=-35,in=110] (1.2,-0.7);

      \draw[black] (-1.5,-1.5) node {$(c)$};
      \draw[anchor=east] (2,-1.65) node {\scriptsize{$t=16.58$}};
    \end{tikzpicture} \\
    \begin{tikzpicture}[>=stealth]
      \draw (0,0) node[inner sep=0] {\includegraphics[height=0.27\textwidth]{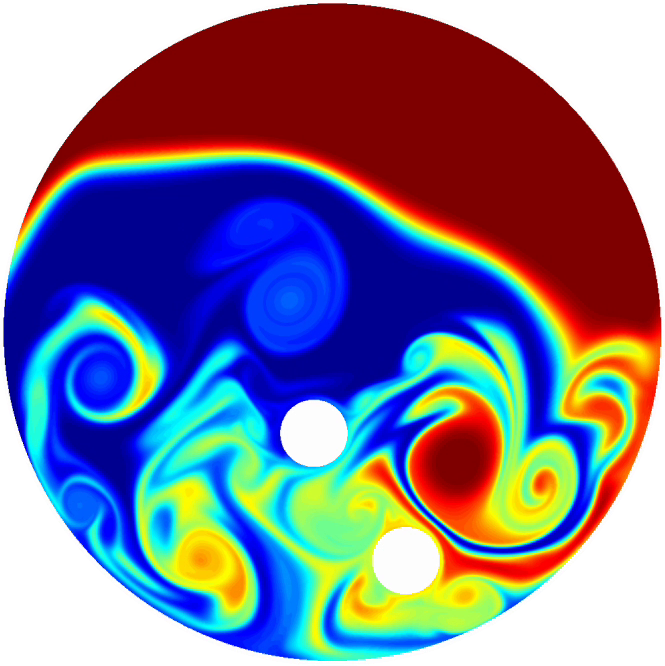}};

      \draw[white,thick,densely dashed,->] (1.1,-1) to [out=-135,in=60] (0.9,-1.3) to [out=-120,in=30] (0.5,-1.6);

      \draw[black] (-1.5,-1.5) node {$(d)$};
      \draw[anchor=east] (2,-1.65) node {\scriptsize{$t=23.25$}};
    \end{tikzpicture} & \hspace{0truecm}
    \begin{tikzpicture}
      \draw (0,0) node[inner sep=0] {\includegraphics[height=0.27\textwidth]{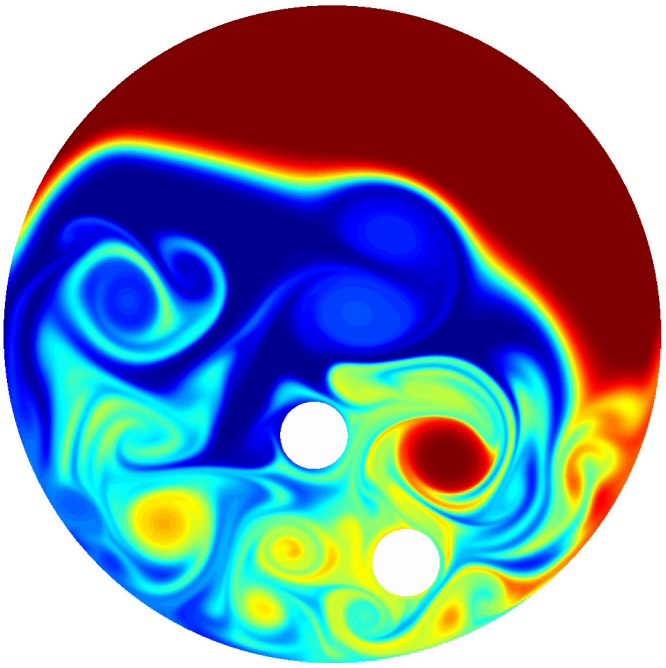}};
      \draw[black] (-1.5,-1.5) node {$(e)$};
      \draw[anchor=east] (2,-1.65) node {\scriptsize{$t=27.83$}};
    \end{tikzpicture} & \hspace{0truecm}
    \begin{tikzpicture}
      \draw (0,0) node[inner sep=0] {\includegraphics[height=0.27\textwidth]{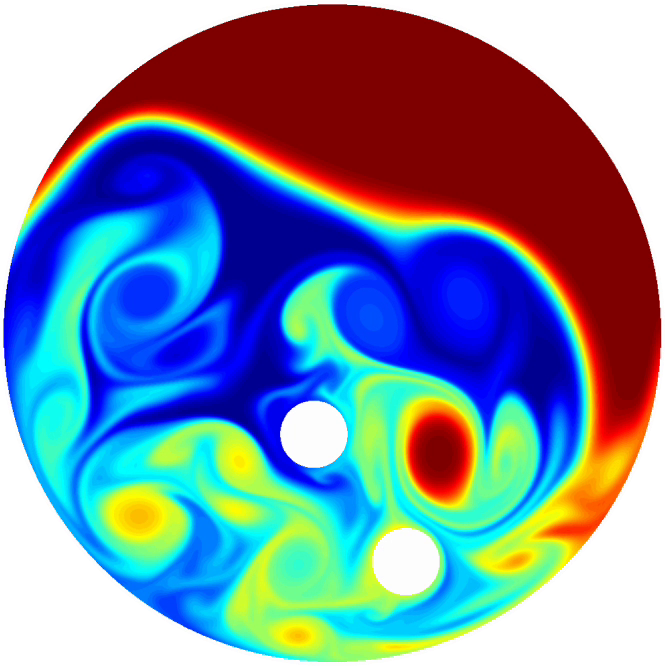}};
      \draw[black] (-1.5,-1.5) node {$(f)$};
      \draw[anchor=east] (2,-1.65) node {\scriptsize{$t=32$}};
    \end{tikzpicture}
  \end{tabular}
  \caption{\label{fig:VelConstraint8} Same as
    figure~\ref{fig:VelConstraint1}, but with an extended control
    window of $T_{\rm{control}}=8.$}
\end{figure}
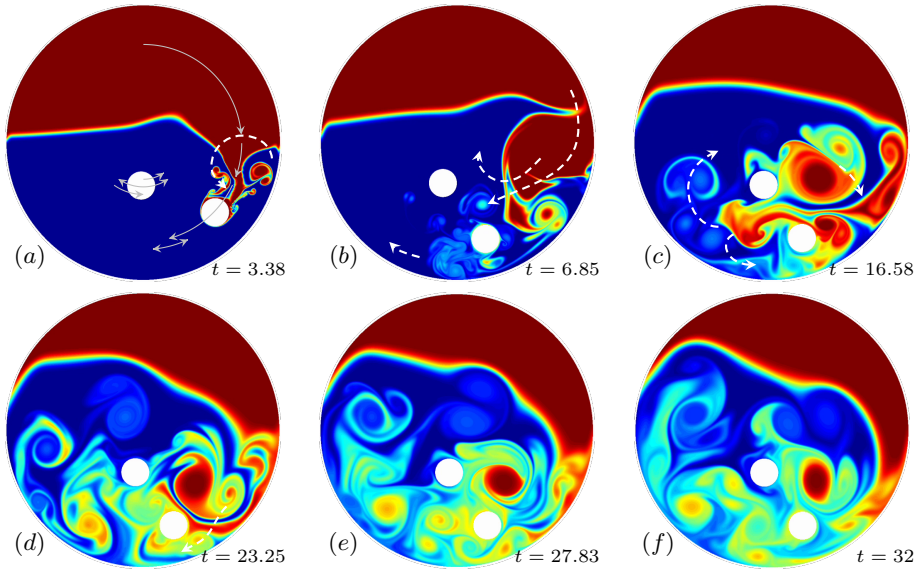

Extending the horizon $T_{\rm{control}}$ over which control is applied
results in a change of strategy (see
figure~\ref{fig:VelConstraint8}). The top stirrer now plunges through
the interface -- but not before stopping and starting on its circular
path towards it. This uneven motion creates more vortical structures
in the stirrer's wake that add to the sole plunging action of the
stirrer itself. The result is a far more distorted interface (and
consequently a lower mix-norm) than would be generated by a simple
traversal. At the end of the motion, a back-and-forth motion is
performed to generate, within the chosen energy and velocity
constraints, additional shed vortices that further interact with the
interface and other vortical elements. The second stirrer does not
follow the strategy of the first. It engages in an oscillatory motion
along its circular path and generates, as before, the resulting
start-and-stop vortices that distort the interface and interact with
the other vortices inside the container. Again, obstruction by the
cylinders (see figures~\ref{fig:VelConstraint8}{\it{b,d}}) and vortex and
wall collisions (see figures~\ref{fig:VelConstraint8}{\it{c,d}}) contribute
to the continued mixing.

In both cases, a gentler stirring strategy is observed. However, the
problem of converging towards a realistic mixing protocol has not been
solved completely. While we explicitly avoid highly localized action
of the stirrers with excessive velocities, we now tend towards
favoring strategies with excessive acceleration. In other words,
within our efforts to limit the total expended energy of the stirrers
while capping their velocities, the optimisation algorithms tends
towards strategies that are characterized by large accelerations (high
velocity gradients). This should not come as a surprise as the
strength of shed vortices from the stirrers' unsteady motion is
proportional to their acceleration. Our imposed constraints do account
for energy and velocities, but not velocity gradients, of the
stirrers. As a consequence, we can seed our binary mixture with
vortical elements of nearly unlimited strength. Again, this divergence
is related to the above-mentioned semi-norm problem: the velocity
field of the binary mixture is not accounted for in the mix-norm, and
thus the optimisation scheme can achieve high-energy {\it{fluid}}
states by highly accelerating stirrers (even though the stirrers'
energy and velocities are capped). To limit the velocity of the
{\it{fluid}}, we have to limit the acceleration of the
stirrers. Again, additional constraints are necessary.

\subsection{Cases 5 and 6: optimisation under energy, velocity and acceleration constraints}

For accomplishing enhanced mixing in binary fluids, the direct-adjoint
optimisation technique makes heavy use of an acceleration-based
strategy: shed vortices generated by the abrupt motion of the stirring
cylinders are injected into both fluids, and their interactions with
the interface, themselves and the container wall yield a low
mix-norm. A limit on this acceleration will result in a limit on the
velocities in either fluid component and thus provide the necessary
restriction for a successful semi-norm optimisation. To this end, we
augment our optimisation scheme by additional terms accounting for the
stirrers' acceleration. This type of penalisation is common in
deblurring of images where strong gradients are detected and
encouraged. In our case, additional projections are used to enforce
the acceleration constraints.

\begin{figure}
  \centering
  \begin{tabular}{ccc}
    \begin{tikzpicture}[>=stealth]
      \draw (0,0) node[inner sep=0] {\includegraphics[height=0.27\textwidth]{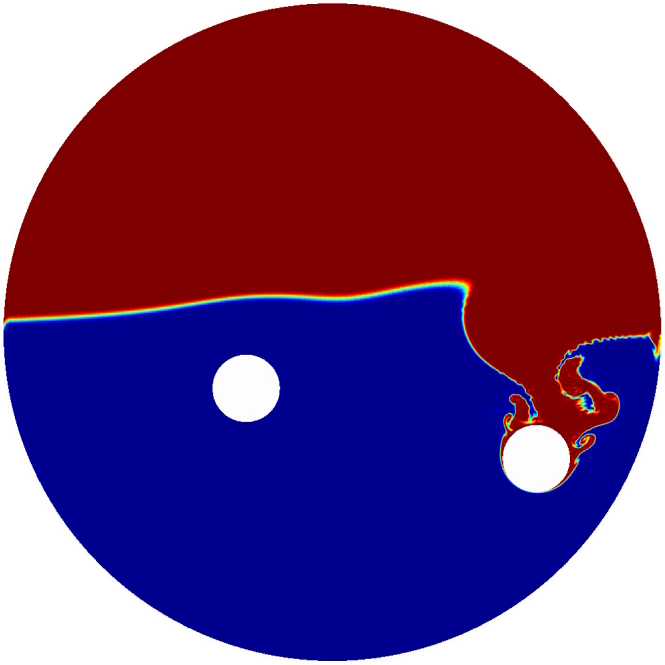}};

     \draw [gray!50,thin,domain=90:-60,samples=100,->] plot ({1.3*cos(\x)}, {1.3*sin(\x)});
     \draw [gray!50,thin,domain=-90:-150,samples=100,->] plot ({0.575*cos(\x)}, {0.575*sin(\x)});

      \draw[white,thick,densely dashed,->] (0.8,0.2) to [out=0,in=90] (1.6,-0.2) to [out=-90,in=50] (1.5,-0.4);
      \draw[white,thick,densely dashed,->] (1.3,-0.2) to [out=-155,in=0] (0.5,-0.5) to [out=180,in=-55] (0.1,-0.2);

      \draw[black] (-1.5,-1.5) node {$(a)$};
      \draw[anchor=east] (2,-1.65) node {\scriptsize{$t=0.68$}};
    \end{tikzpicture} & \hspace{0truecm}
    \begin{tikzpicture}[>=stealth]
      \draw (0,0) node[inner sep=0] {\includegraphics[height=0.27\textwidth]{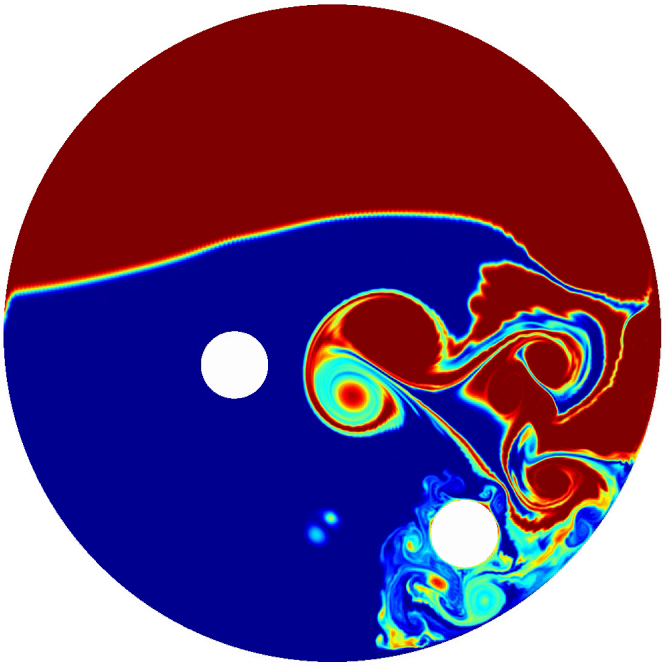}};

      \draw[white,thick,densely dashed,->] (-0.1,0.1) to [out=135,in=-90] (-0.3,0.5) to [out=90,in=-155] (0.5,1.3);
      \draw[white,thick,densely dashed,->] (1.2,-0.2) to [out=-110,in=0] (0.7,-0.7) to [out=180,in=-110] (0.4,-0.1);

      \draw[black] (-1.5,-1.5) node {$(b)$};
      \draw[anchor=east] (2,-1.65) node {\scriptsize{$t=1.58$}};
    \end{tikzpicture} & \hspace{0truecm}
    \begin{tikzpicture}[>=stealth]
      \draw (0,0) node[inner sep=0] {\includegraphics[height=0.27\textwidth]{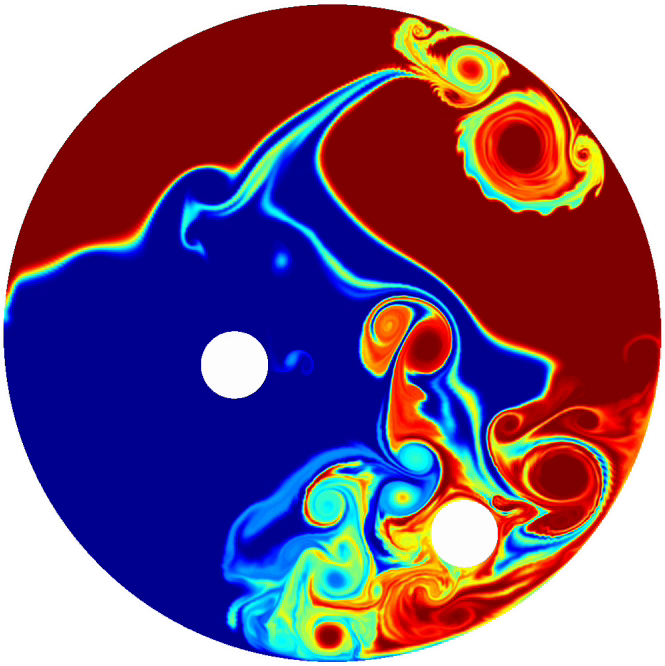}};

      \draw[white,thick,densely dashed,->] (0.5,0.2) -- (0.7,0.45);
      \draw[white,thick,densely dashed,->] (0.9,0.7) -- (0.7,0.45);
      \draw[white,thick,densely dashed,->] (0.6,0.55) to [out=135,in=-90] (0.5,0.75) to [out=90,in=-155] (0.7,0.9);
      \draw[white,thick,densely dashed,->] (0.8,0.35) to [out=-45,in=0] (0.5,-0.2) to [out=180,in=-90] (0.1,0.4) to [out=90,in=-115] (0.3,1);
      \draw[white,thick,densely dashed,->] (-0.8,-0.4) to [out=180,in=-65] (-1.4,0);

      \draw[black] (-1.5,-1.5) node {$(c)$};
      \draw[anchor=east] (2,-1.65) node {\scriptsize{$t=2.74$}};
    \end{tikzpicture} \\
    \begin{tikzpicture}[>=stealth]
      \draw (0,0) node[inner sep=0] {\includegraphics[height=0.27\textwidth]{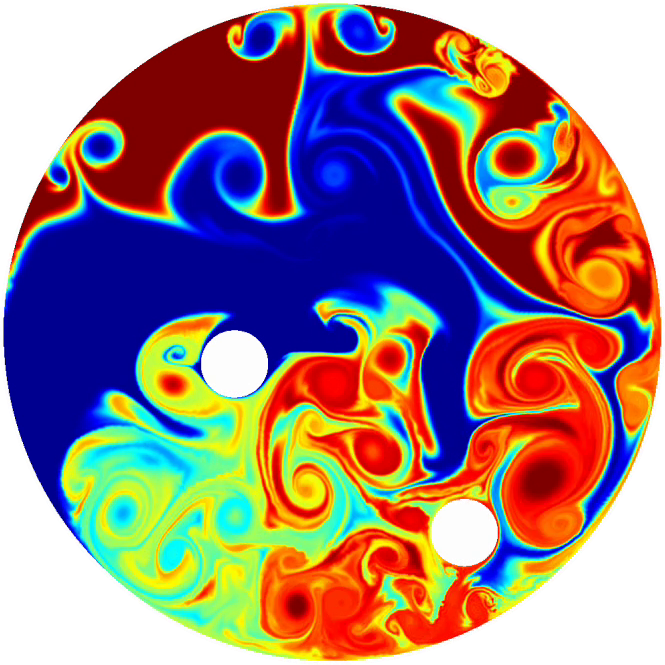}};

      \draw[white,thick,densely dashed,->] (1.3,-0.5) to [out=45,in=-90] (1.6,0.2) to [out=90,in=45] (0.3,-0.1);
      \draw[white,thick,densely dashed,->] (-1.1,-0.2) to [out=180,in=180] (-1.1,-0.6) to [out=0,in=-100] (-0.8,0.4) to [out=80,in=-125] (-0.5,1);

      \draw[black] (-1.5,-1.5) node {$(d)$};
      \draw[anchor=east] (2,-1.65) node {\scriptsize{$t=5.43$}};
    \end{tikzpicture} & \hspace{0truecm}
    \begin{tikzpicture}[>=stealth]
      \draw (0,0) node[inner sep=0] {\includegraphics[height=0.27\textwidth]{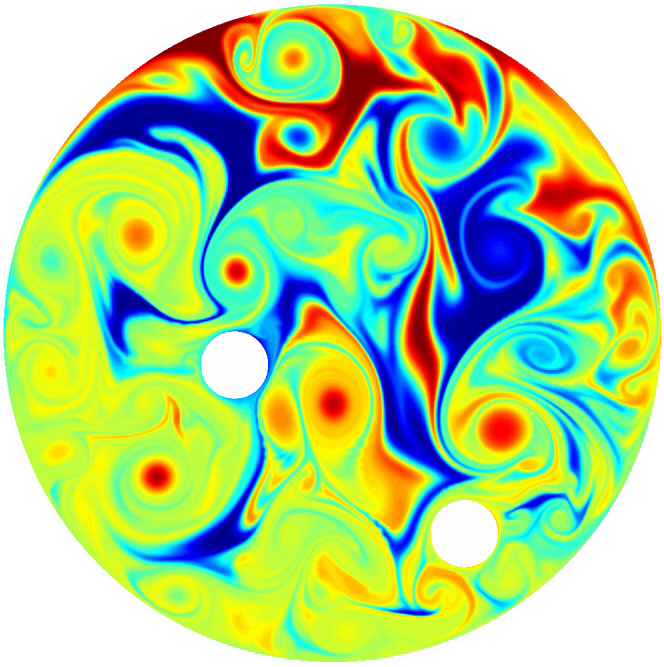}};

      \draw[white,thick,densely dashed,->] (0.2,-0.3) to [out=0,in=155] (0.5,-0.45);
      \draw[white,thick,densely dashed,->] (0.8,-0.3) to [out=180,in=25] (0.5,-0.45);
      \draw[white,thick,densely dashed,->] (0.5,-0.5) to [out=-90,in=5] (-0.4,-1);
      \draw[white,thick,densely dashed,->] (0.2,1.2) to [out=0,in=-150] (0.8,1.35);

      \draw[black] (-1.5,-1.5) node {$(e)$};
      \draw[anchor=east] (2,-1.65) node {\scriptsize{$t=12.36$}};
    \end{tikzpicture} & \hspace{0truecm}
    \begin{tikzpicture}
      \draw (0,0) node[inner sep=0] {\includegraphics[height=0.27\textwidth]{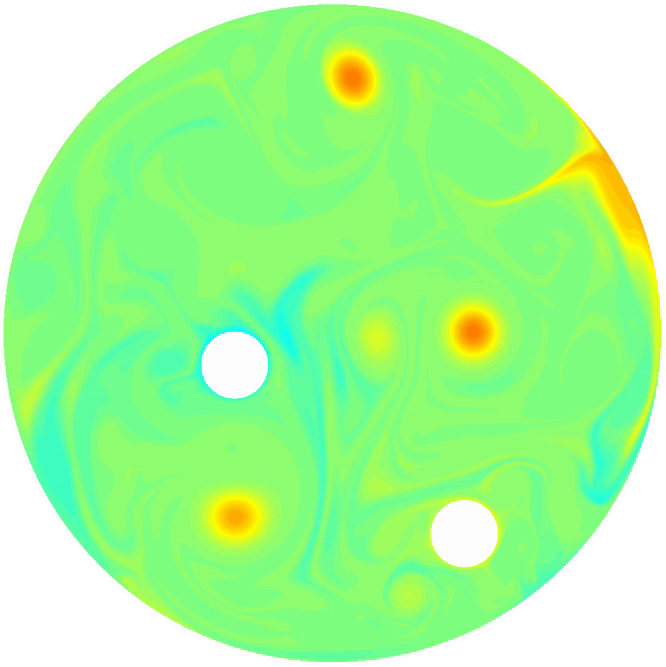}};
      \draw[black] (-1.5,-1.5) node {$(f)$};
      \draw[anchor=east] (2,-1.65) node {\scriptsize{$t=32$}};
    \end{tikzpicture}
  \end{tabular}
  \caption{\label{fig:AccConstraint1} Mixing optimisation based on
    energy, velocity and acceleration constraints for the
    stirrers. The time horizon for applying control is
    $T_{\rm{control}}=1.$ Shown are iso-contours of the passive scalar
    at selected instances. The optimisation algorithm includes
    information over a time window of $T_{\rm{info}}=8.$}
\end{figure}
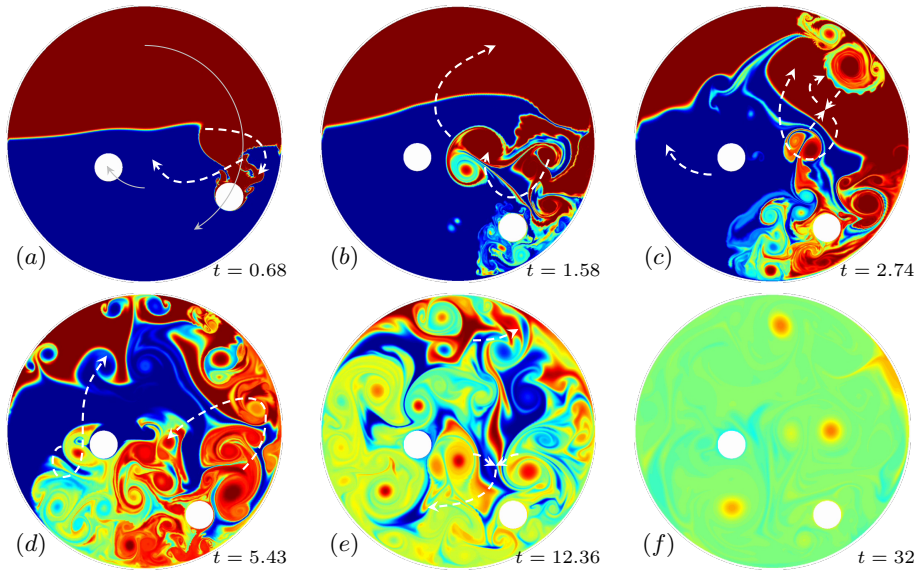

For the short-term control with $T_{\rm{control}}=1$ (see
figure~\ref{fig:AccConstraint1}), the optimal strategy now includes a
plunging of the first cylinder through the interface, while the second
cylinder continues in a straight manner towards the interface but
stops short of it. The wake vortices of the first cylinder, as well as
the (weaker) start-and-stop vortices of both cylinders, are
responsible for the bulk of the mixing. As before, complex vortex
collisions (see figure~\ref{fig:AccConstraint1}{\it{c}}), stirrer
obstruction (see figures~\ref{fig:AccConstraint1}{\it{c,d}}) and wall
interactions (see figure~\ref{fig:AccConstraint1}{\it{d}}) contribute
greatly to the breakdown of scales, the generation of filaments (see
figure~\ref{fig:AccConstraint1}{\it{e}}) and the eventual mixing of the
binary fluid (see figure~\ref{fig:AccConstraint1}{\it{f}}).

A longer control horizon of $T_{\rm{control}}=8$ yields a more varied
stirring protocol. The first cylinder makes a farther excursion,
plunging through the interface (not without stopping to generate
additional shed vortices close to the interface) before stopping close
to the interface and shedding two stop vortices. The second cylinder
first approaches the interface, ejects a stop vortex before reversing
and stopping short of the interface with another stop vortex. The
generated structures interact with themselves and the wall to break
down the binary fluid into a homogeneous mixture, although of less
homogeneity (larger mix-norm) than for the short-term strategy. This
reduced homogeneity can be attributed more to the restricted velocity
range (which is due to the constraint of an equal energy budget across
both time horizons) than the larger time-horizon strategy. The lower
velocity maximum, combined with the limitation on acceleration,
impedes the same amount of vortex shedding than for the shorter
time-horizon case. For this reason, mixing cannot be as efficient.

At no point during either optimisation has the energy, velocity or
acceleration of the stirrers exceeded the specified limits. As a
consequence, these latter strategies are amenable to implementation in
an experimental or industrial setting.

\begin{figure}
  \centering
  \begin{tabular}{ccc}
    \begin{tikzpicture}[>=stealth]
      \draw (0,0) node[inner sep=0] {\includegraphics[height=0.27\textwidth]{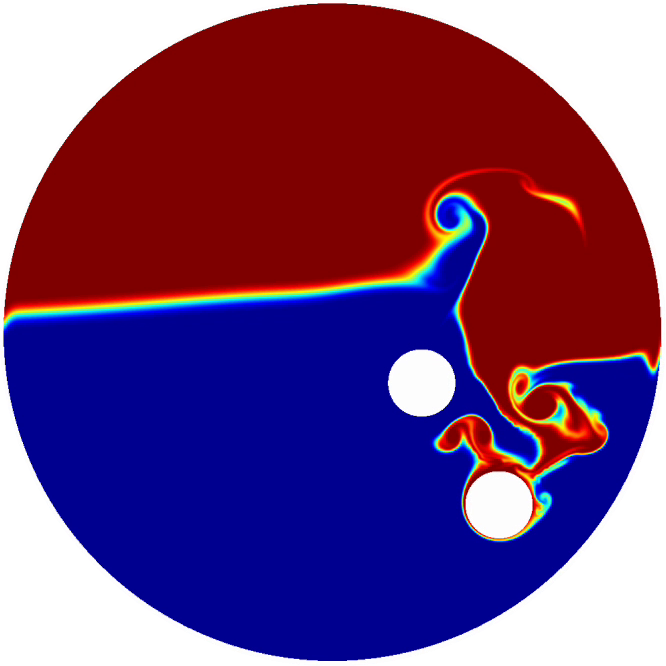}};

     \draw [gray!50,thin,domain=90:45,samples=100,->] plot ({1.3*cos(\x)}, {1.3*sin(\x)});
     \draw [gray!50,thin,domain=40:-10,samples=100,->] plot ({1.3*cos(\x)}, {1.3*sin(\x)});
     \draw [gray!50,thin,domain=-15:-25,samples=100,->] plot ({1.3*cos(\x)}, {1.3*sin(\x)});
     \draw [gray!50,thin,domain=-30:-180,samples=100,->] plot ({1.3*cos(\x)}, {1.3*sin(\x)});

     \draw [gray!50,thin,domain=-90:0,samples=100,->] plot ({0.575*cos(\x)}, {0.575*sin(\x)});
     \draw [gray!50,thin,domain=0:-170,samples=100,->] plot ({0.675*cos(\x)}, {0.675*sin(\x)});

      \draw[white,thick,densely dashed,->] (0.4,0) to [out=35,in=-110] (0.65,0.4);
      \draw[white,thick,densely dashed,->] (0.8,0.1) to [out=10,in=90] (1.2,-0.2);
      \draw[white,thick,densely dashed,->] (0.85,0.75) to [out=45,in=-90] (1,1.1);
      \draw[white,thick,densely dashed,->] (0.85,0.75) to [out=45,in=180] (1.15,0.85) to [out=0,in=95] (1.35,0.4);

      \draw[black] (-1.5,-1.5) node {$(a)$};
      \draw[anchor=east] (2,-1.65) node {\scriptsize{$t=4$}};
    \end{tikzpicture} & \hspace{0truecm}
    \begin{tikzpicture}[>=stealth]
      \draw (0,0) node[inner sep=0] {\includegraphics[height=0.27\textwidth]{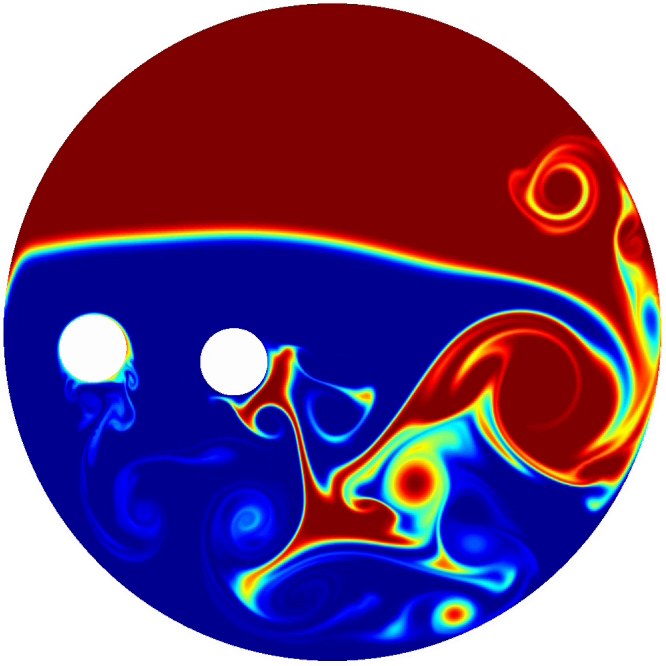}};

      \draw[white,thick,densely dashed,->] (0.8,-1) to [out=45,in=-40] (0.55,-0.55);
      \draw[white,thick,densely dashed,->] (1.15,-0.5) to [out=-135,in=-50] (0.6,-0.5);
      \draw[white,thick,densely dashed,->] (0.4,-0.9) to [out=-125,in=15] (-0.2,-1.3);
      \draw[white,thick,densely dashed,->] (-1.5,0.05) to [out=90,in=-45] (-1.6,0.3);
      \draw[white,thick,densely dashed,->] (-1.1,0.05) to [out=90,in=-60] (-1.2,0.4);

      \draw[black] (-1.5,-1.5) node {$(b)$};
      \draw[anchor=east] (2,-1.65) node {\scriptsize{$t=6.91$}};
    \end{tikzpicture} & \hspace{0truecm}
    \begin{tikzpicture}[>=stealth]
      \draw (0,0) node[inner sep=0] {\includegraphics[height=0.27\textwidth]{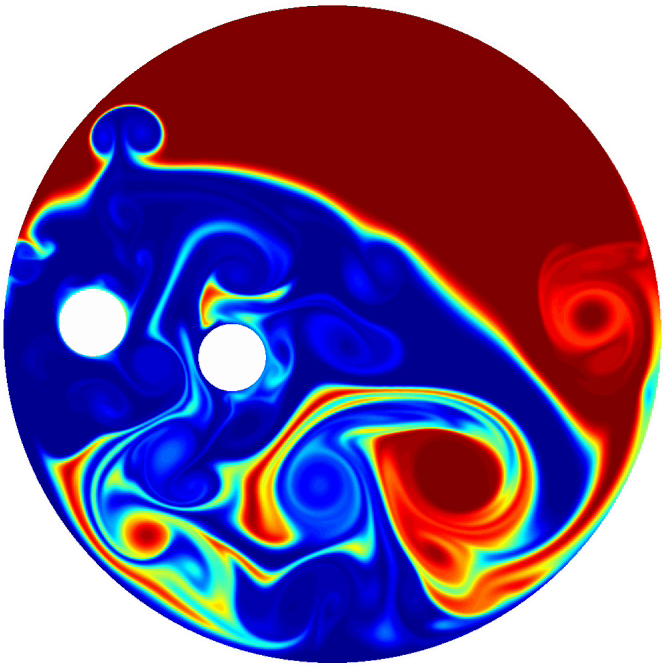}};

      \draw[white,thick,densely dashed,->] (0.5,-0.8) to [out=160,in=-160] (0.4,-0.1);
      \draw[white,thick,densely dashed,->] (-0.85,0.55) to [out=60,in=-165] (0.4,1.3);

      \draw[black] (-1.5,-1.5) node {$(c)$};
      \draw[anchor=east] (2,-1.65) node {\scriptsize{$t=9.82$}};
    \end{tikzpicture} \\
    \begin{tikzpicture}[>=stealth]
      \draw (0,0) node[inner sep=0] {\includegraphics[height=0.27\textwidth]{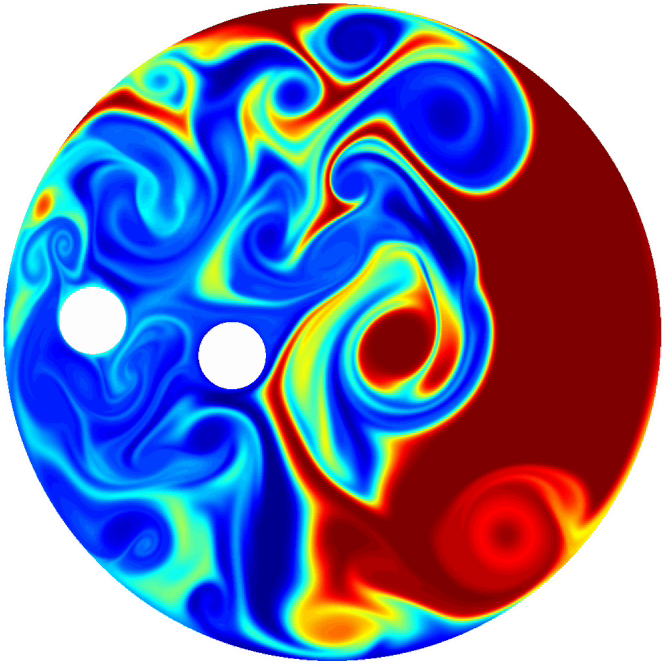}};

      \draw[white,thick,densely dashed,->] (0.6,-0.2) to [out=-20,in=100] (0.9,-0.8);
      \draw[white,thick,densely dashed,->] (0.9,0.7) to [out=-60,in=80] (1,-0.6);
      \draw[white,thick,densely dashed,->] (0.5,-1.2) to [out=180,in=-75] (-0.4,-0.5);

      \draw[black] (-1.5,-1.5) node {$(d)$};
      \draw[anchor=east] (2,-1.65) node {\scriptsize{$t=17.6$}};
    \end{tikzpicture} & \hspace{0truecm}
    \begin{tikzpicture}
      \draw (0,0) node[inner sep=0] {\includegraphics[height=0.27\textwidth]{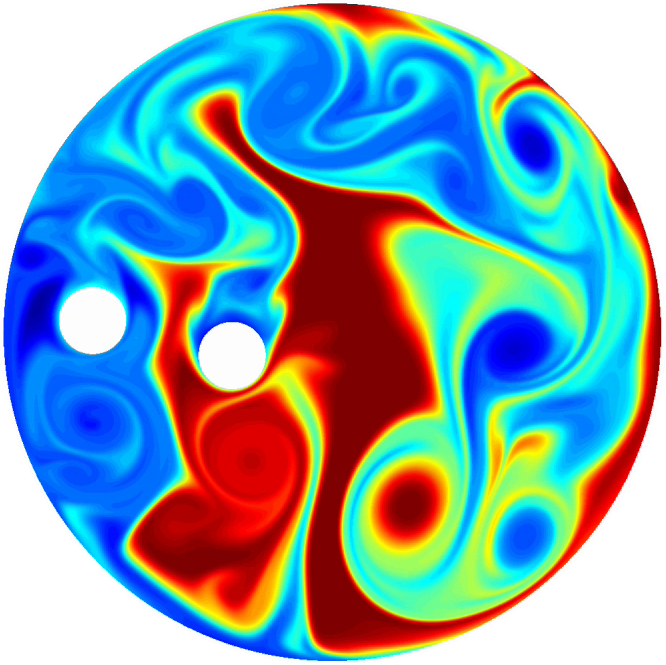}};
      \draw[black] (-1.5,-1.5) node {$(e)$};
      \draw[anchor=east] (2,-1.65) node {\scriptsize{$t=27.64$}};
    \end{tikzpicture} & \hspace{0truecm}
    \begin{tikzpicture}
      \draw (0,0) node[inner sep=0] {\includegraphics[height=0.27\textwidth]{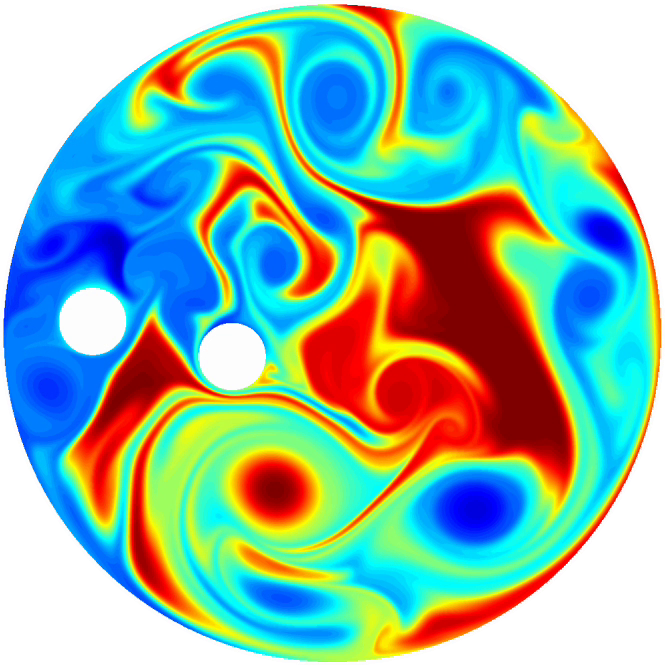}};


      \draw[black] (-1.5,-1.5) node {$(f)$};
      \draw[anchor=east] (2,-1.65) node {\scriptsize{$t=32$}};
    \end{tikzpicture}
  \end{tabular}
  \caption{\label{fig:AccConstraint8} Same as
    figure~\ref{fig:AccConstraint1}, but with an extended control
    window of $T_{\rm{control}}=8.$}
\end{figure}
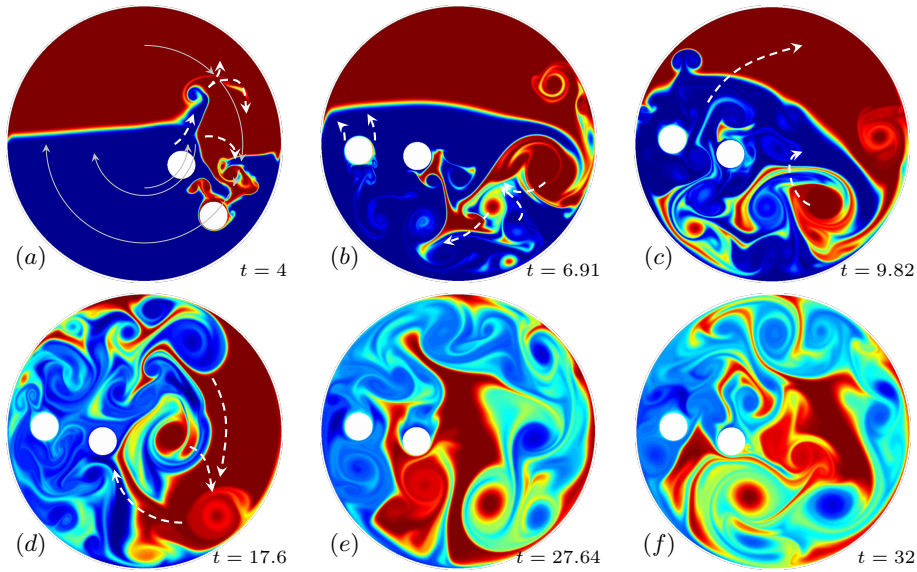

\section{Summary, conclusions and remaining challenges}

A direct-adjoint optimisation methodology has been applied to the
problem of mixing of binary fluids. A circular configuration with two
embedded stirrers on circular paths has been chosen, and the
velocities of these two stirrers over a user-specified time interval
have been determined in an attempt to enhance the homogeneity of the
binary mixture. The gradient-based optimisation is effective in
finding stirring protocols that yield enhanced mixing results, but
convergence towards an optimum crucially relies on imposing proper
constraints on the iterative algorithm. Since the mixing efficiency is
based on only one dynamic variable, the passive scalar $\theta,$ but
disregards the velocity fields, additional external constraints have
to be imposed to properly define a feasible optimum. These constraints
have to enforce limitations on the encountered fluid velocities, which
are forced by accelerating the stirrers. The accelerations, in turn,
inject vortical structures into both fluids by unsteady,
Stokes-layer-type shedding of vortices. Thus, by restricting the
stirrers' maximum acceleration, we arrive at a properly stated
optimisation problem and a convergent direct-adjoint algorithm.

Under these conditions, the optimal strategy that improves on the
mixedness of the binary fluid utilises a combination of prototypical
mixing techniques, consisting of plunging, unsteady vortex shedding,
collisions between vortices and the wall, and obstructions by the
stirrers. While unsteady vortex shedding is the key strategy for the
unconstrained (or insufficiently constrained) case, a more balanced
protocol ensues when excessive accelerations are increasingly
penalised. Nonetheless, a rather counterintuitive optimal mixing
strategy has been determined for short and longer time-horizons. As a
general tendency, shorter control windows reach a lower mix-norm
state, as the stirrer motion is more vigorous over a more limited
horizon.

In conclusion, the above direct-adjoint approach to mixing of binary
fluids -- when combined with an efficient spectral simulation scheme,
a Brinkman-type penalisation to accommodate moving bodies, and a
systematic checkpointing technology -- has proven an effective and
robust tool to design stirrer strategies for the enhancement of
mixing. After proper limitations on the stirrers' acceleration have
been taken into account and a proper measure of mixedness has been
defined, stirring strategies exploiting the full range of fluid
processes induced by fluid-structure interactions are found, that
suggest realisable modifications to commonly employed stirrer-induced
mixing methods for industrial applications. The accomplished increase
in mixing efficiency is summarised in table~\ref{tab:Cases} which
lists the mix-norm values for $t=1$, $t=8$ and $t=32.$ With an imposed
acceleration penalisation, the short control horizon
($T_{\rm{control}}=1$) gives markedly better results than a longer one
($T_{\rm{control}}=8$), even though the difference is less pronounced
after rest-inertia and diffusion set in. The number of iterations
taken by the direct-adjoint algorithm is displayed as well; longer
time horizons typically converge faster, owing to a less abrupt
stirrer protocol. The short-horizon case with only energy penalisation
(case 1, above) is included for comparison.

\begin{table}
  \centering
  \begin{tabular}{cccccccc}
    &\phantom{1}& $T_{control}$ & iterations &\phantom{1}& $\Vert \theta \Vert_{mix,t=1}$ &
    $\Vert \theta \Vert_{mix,t=8}$ & $\Vert \theta \Vert_{mix,t=32}$ \\
    && & && & & \\
    AccPen && $T_{\rm{control}}=1$ & 11 && 0.3230 & 0.0745 & 0.0433 \\
    AccPen && $T_{\rm{control}}=8$ & 5 && 0.3885 & 0.2437 & 0.0558 \\
    && & & & & \\
    EnPen && $T_{\rm{control}}=1$ & 12 && 0.3769 & 0.1724 & 0.0597
  \end{tabular}
  \caption{\label{tab:Cases} Summary of results for acceleration
    penalisation (AccPen) and energy-only penalisation (EnPen). Short
    ($T_{\rm{control}}=1$) and longer ($T_{\rm{control}}=8$) control
    horizons are displayed, together with the number of iterations
    taken by the direct-adjoint optimisation algorithm.}
\end{table}

Despite a successful increase of mixing efficiency via uncommon and
unexpected strategies and despite corroborating and supporting the
chosen computational approach, the present study also raises a number
of challenges and shortcomings. Foremost among them is the fact that
gradient-based optimisation with nonlinear partial differential
equations as constraints can only assure a local optimum; a globally
optimal solution requires an additional, and often prohibitive,
methodology, such as simulated annealing or other variations of the
same concept. While a global optimum may certainly be desirable, we
point out that the improvements in mixing efficiency from a local
solution (as shown in this study) would already have a respectable
impact on mixing results due to its omnipresence in many industrial
settings. In this sense, improvements suggested by locally optimal
solutions would be most welcome.

A further challenge consists of additional constraint handling,
imposed by mechanical restrictions on the stirrer motion. Penalisation
methods or auxiliary projections imposed on the gradient information
are conceivable to address this issue.

The length of the optimisation window $T_{\rm{info}}$ places a bound
on the overall algorithm. The adjoint part of the simulations computes
the sensitivity of the output functional (mix norm) with respect to
our control parameters (stirrer velocities along their paths). For
increasingly large optimisation windows these sensitivities diverge
due to the quasi-chaotic behavior of the direct problem. As a result,
meaningful sensitivity to aid our optimisation will get overwhelmed by
general sensitivity due to chaotic motion, and if meaningful
sensitivity is lost, the optimisation algorithm will stagnate or even
diverge. An advance to higher Reynolds or P\'eclet numbers will
encounter similar issues. While techniques to overcome this
predicament are currently being developed~\citep{Blonigan}, their
cost-efficient application to complex systems, such as mixing, is
still an open problem.

However, within the constraints of this study, with a physical problem
this rich in possibilities and with a computational approach to match,
there is an abundance of extensions and opportunities. Besides obvious
explorations of other parameter combinations, the optimisation of the
stirrers' shape is certainly within the capabilities of the
computational framework; a preliminary study in this direction can be
found in~\cite{Eggl2019} using cycloids and trochoids as
cross-sectional stirrer geometries. The path of the stirrers (in our
study, concentric circles) can also be optimised; a collision-avoiding
constraint may pose an additional challenge in this case. Furthermore,
since the wall constitutes an important component in the breakup of
vortices, an optimisation of wall motion or wall corrugation may
further facilitate a more rapid breakdown in scales. With a view
towards industrial applications, a non-Newtonian fluid model may be
implemented. Finally, injection mixing (where the unmixed fluids are
introduced into a mixing device or passive baffle system, and
extracted when fully mixed) could be treated within the same
direct-adjoint framework. These possible extensions, some of which
will be reported in future efforts, attest to the flexibility and
efficacy of the computational setup; marked enhancements in mixing are
expected in the above cases.

\vspace{1.0cm}
\noindent Declaration of Interests. The authors report no conflict of interest.
\bibliographystyle{jfm}
\bibliography{jfm_mixingNEW}

\end{document}